\documentclass[twocolumn]{aastex62}

\usepackage{graphicx} 
\usepackage{xcolor}
\usepackage{mathtools}
\usepackage{amsmath}
\usepackage{lineno}

\shorttitle{Simulated Observations of Multiphase Galactic Winds}
\shortauthors{Lita M. de la Cruz, Evan E. Schneider, Eve C. Ostriker}

\begin{document}

\title{Synthetic Absorption Lines from Simulations of Multiphase Gas in Galactic Winds}

\correspondingauthor{Lita M. de la Cruz}
\email{l.delacruz@princeton.edu}

\author{Lita M. de la Cruz}
\affil{Department of Astrophysical Sciences, Princeton University, 4 Ivy Lane, Princeton, NJ 08540, USA}
\author[0000-0001-9735-7484]{Evan E. Schneider}
\affil{Department of Physics and Astronomy, University of Pittsburgh, 3941 O'Hara Street, Pittsburgh, PA 15260}
\author[0000-0002-0509-9113]{Eve C. Ostriker}
\affil{Department of Astrophysical Sciences, Princeton University, 4 Ivy Lane, Princeton, NJ 08540, USA}

\begin{abstract}
Supernova-driven galactic winds are multiphase streams of gas that are often observed flowing at a range of velocities out of star-forming regions in galaxies. In this study, we use high resolution 3D simulations of multiphase galactic winds modeled with the hydrodynamics code \textit{Cholla} to investigate the connection between numerical studies and observations. Using a simulated interaction between a hot $T\sim 10^{7}\,\rm K$ supernova-driven wind and a cool $T\sim 10^{4}\,\rm K$ cloud of interstellar material, we create mock observables, including the optical depth and covering fraction of six commonly observed ions (Si II, C II, Si IV, C IV, N V, and O VI) as a function of gas velocity. We compare our mock observables to surveys of galactic winds in the literature, finding good agreement with velocities and profiles of the low ions. We then compute ``empirical" values for the optical depth and covering fraction following observational techniques, and compare them to the values calculated directly from the simulation data. We find that the empirically computed values tend to underestimate the ``true" value of $\tau$ for ions with high optical depth, and overestimate the ``true" value of $\tau$ for ions with low optical depth, relative to the simulated data.
\end{abstract}

\keywords{galaxies: evolution - starburst - hydrodynamics - ISM: outflows - supernovae}

\section{Introduction}

In 1963, an optical emission line spectrum revealed high velocity outflows of cool ionized gas in the central region of the nearby irregular galaxy M82 \citep{Lynds..et..al..1963}. The large velocities indicated an outburst powered by a massive amount of energy - a phenomenon that has since captivated the interest of researchers for almost 60 years. Subsequent observations have revealed that outflows are common in star-forming galaxies, driven by the momentum and energy input of massive stars in the form of supernovae, stellar winds, cosmic rays, and radiation \cite[see][for reviews]{Heckman..et..al..1993,Veilleux..et..al..2005, Heckman..et..al..2017,Veilleux..et..al..2020}. 

Outflows are now understood to play an important role in the evolution of galaxies and the intergalactic medium (IGM). By controlling the amount of gas in galaxies available to convert into stars, winds limit the baryon content of galaxies
\citep{Larson..1974, Somerville..et..al..1999, Keres..et..al..2005, Crain..et..al..2009, Dave..et..al..2011, Moster..et..al..2013, Behroozi..et..al..2013, Genel..et..al..2014, Muratov..et..al..2015}. By transporting metals out of star-forming regions, outflows regulate the metallicity of stars and gas in galaxies, setting the mass-metallicity relation \citep{Tremonti..et..al..2004}. This same process enriches the circumgalactic medium (CGM) and intergalactic medium (IGM) with metals \citep{Simcoe..et..al..2004, Steidel10, Werk13, Rubin..et..al..2014, Hopkins..et..al..2014}. Thus, understanding the process by which outflows remove gas from galaxies is key to understanding galaxy evolution. 

Supernovae-driven winds are multiphase, and have been observed at a broad range of densities and temperatures \citep{Heckman..et..al..1990, Dave..et..al..1998, Martin..2005, Strickland..et..al..2009, Erb..et..al..2012, Rubin..et..al..2014}. 
Generating a complete picture of  multiphase winds,
however, 
requires great time and effort and 
the use of many different observational instruments. With the high spatial resolution of the Chandra X-ray Observatory, the hot $10^{7}\,\rm K$, supersonic wind created by supernova explosions can be observed directly, but only for the closest galaxies \citep{Griffiths..et..al..2000, Strickland..et..al..2007}.
This phase in any case carries very little of the mass that eventually leaves galaxies, in spite of its role as a primary driver; only by interacting with denser ISM phases is the hot, fast wind able to create a massive outflow.   

The cooler ($T < 10^6\,\mathrm{K}$), indirectly accelerated gas in outflows is commonly studied by observers because of the many strong absorption and emission lines \citep{Heckman..et..al..2000}. Given the $n^2$ dependence on density, detecting outflows in emission lines of distant galaxies tends to be challenging, so outflows in these sources are more typically studied via absorption lines. Measurements of absorption line profiles give direct insight into the velocity distribution and column density of the outflowing gas, as well as indirect measurements of metallicity, optical depth, covering fraction, and mass outflow rates. Commmonly detected species in absorption include C, N, O, Mg, Ca, Si, S, and Fe. These species in their various ionization states probe a range of gas temperatures ranging from cold neutral gas ($T\sim 10^{3}\,\mathrm{K}$) to hot ionized gas ($T \sim 10^{6}\,\mathrm{K}$).

The installation of the Cosmic Origins Spectrograph (COS) on the Hubble Space Telescope (HST) significantly increased the efficiency of observations of this neutral and ionized gas phase in the outskirts of nearby UV bright starburst galaxies \citep{Shull..2009, Green..et..al..2012, Claus..et..al..2013, Wood..et..al..2015, Chisholm..et..al..2015, Chisholm..et..al..2016a,  Chisholm..et..al..2016b, Chisholm..et..al..2017}. \cite{Tumlinson..et..al..2011} and \cite{Thom..et..al..2011} used COS to study the gaseous halos of low redshift galaxies and examined the metallicity and ionization of O VI absorption lines that surrounded these halos. In a particularly detailed observation of a gravitationally-lensed galaxy, \citet{Chisholm..et..al..2018} studied the O VI doublet ($\lambda$ = 1032$\mathring{ \rm A}$ ), one of the few observable tracers that is provisionally created during mass-loading (a mixing phase between the cool and hot gas). That study found that the O VI doublet is a unique species that has two regimes: at high velocity it mimics low-ionization optically thin lines, and at low velocity it mimics strong saturated lines. In many cases, the creation of this species is ambiguous, and it can be found in many different environments such as galaxies, the CGM, filaments, and the IGM \citep{Savage..et..al..1998, Trip..et..al..2001, Lehner..et..al..2009, Werk16}. 

There are many observational studies of the warm neutral and (photo)ionized  $10^{4} \rm K$  gas in outflows, because this phase has strong absorption and emission lines in the rest-frame ultraviolet and optical \citep{Heckman..et..al..2000}. Early optical emission line studies of galactic outflows used narrow-band images and long-slit spectroscopy of nearly edge-on bright IR galaxies \citep{ McCarthy..et..al..1987, Heckman..et..al..1990, Lehnert..et..al..1996}. Other observational work uses absorption lines from the Na I ``D" doublet in far-IR starburst galaxies \citep{Heckman..et..al..2000, Martin..2005, Rupke..et..al..2005b}. Due to its low ionization potential of $5.1 \rm eV$, ground state Na I D can effectively probe the neutral gas phase ($T\lesssim 8000 \rm K$) in galactic winds. The ubiquity of this doublet in ultraluminous IR galaxies (ULIRGs) makes it a regularly used tracer in studies of these systems. However, this delicate species is shielded from photoionization by the immense amount of dust found in ULIRGS \citep{Heckman..et..al..2000, Heckman..et..al..2001, Martin..2006, Heckman..et..al..2017}, and is easily destroyed in less dusty systems. UV absorption lines that are redshifted into the optical, such as Mg II and Fe II, can be used to study outflows 
of gas at $T\sim 10^{4}\,\rm K$ in higher redshift galaxies via stacking of spectra \citep{Weiner..et..al..2009, Maltby2019, Sugahara2019}. These species can survive the strong UV radiation fields, but require high spectral resolution and individual observations generally require bright galaxies to achieve adequate signal-to-noise \citep{Kornei..et..al..2012, Rubin..et..al..2014}. In some unique cases, such as gravitationally lensed systems, these species can be studied in detail in high-redshift galaxies with lower star formation rates \citep{Chisholm..et..al..2018}.

%
%Moreover, all of these studies focus on the neutral or low ionization gas phases of galactic winds and 
%don't 
%do not directly trace the hot $10^{7}$ K gas that powers the outflow. 
%By using the high spatial resolution of the Chandra X-ray Observatory, the hot supersonic fluid created by supernova explosions can be observed directly, but only for the closest galaxies \citep{Griffiths..et..al..2000, Strickland..et..al..2007}.
%Thus, generating a complete image of a multiphase wind for any given galaxy requires great time and effort and the use of many different instruments.

%\cite{Chisholm..et..al..2018} studies the O VI doublet ($\lambda$ = 1032$\mathring{A}$ ), one of the few observable tracers that is provisionally created during mass-loading.These spectroscopic observations study absorption lines from a gravitationally lensed high red-shift galaxy with a signal-to-noise (SNR) close to that of the O VI 1032$\mathring{A}$ absorption line.

%The changing properties of cold and warm outflows of these multiphase winds was studied through spectroscopic observations in near star-forming regions (\cite{Leroy..et..al..2015}).
%In addition, the efficiency of photoionized outflows through the use of the Sobolov approximation and elaborate photoionization models was also studied \cite{Chisholm..et..al..2017}.

On the theoretical side too, much effort has been spent trying to understand the physics of galactic outflows. However, understanding the different processes that govern the different phases of gas has proven to be challenging. Early analytic work by \citet{Chevalier..and..Clegg..1985} provided a model of the observed X-ray emission produced by the directly-thermalized hot plasma in supernova-driven outflows. However, these spatially-resolved X-ray observations are limited to nearby galaxies, and are not easy to link to observations of other phases. In order to produce a complete theoretical picture, simulations must compare to the many down-the-barrel absorption line measurements  used to study the lower-temperature gas in both local and more distant galaxies. But this approach too has challenges. Converting these absorption-line measurements into direct physical properties like mass outflow rates often requires a host of limiting assumptions. Likewise, converting simulated data into directly-observed quantities is non-trivial.

Numerical simulations can provide an avenue to better link our theoretical and observational models of outflows. In recent years, much theoretical work has focused specifically on understanding the physics that governs the interaction between a hot, supernova-driven wind and embedded cool clouds of material. In a seminal numerical study, \cite{Klein..et..al..1994} formulated a characteristic timescale that describes the cloud-shock interaction: the cloud-crushing time $t_{\rm cc}$. The cloud-crushing time roughly refers to the amount of time that it takes the initial shock to propagate through the cloud. Analyzing 2D adiabatic simulations, \cite{Klein..et..al..1994} used this timescale to describe the various stages of cloud evolution. The authors concluded that on large scales, spherical clouds in a background wind are destroyed in only a few $t_{\rm cc}$. Subsequently, the cloud material mixes efficiently at $t\approx$ 4-5$t_{\rm cc}$ with the surrounding medium via small scale instabilities.

In subsequent years, different studies have focused on the different physical mechanisms that may affect this interaction, including magnetic fields \citep{MacLow..et..al..1994, Gregori2000, Banda-Barragan2016}, radiative cooling \citep{Fragile2005, Cooper..et..al..2009, Scannapieco..et..al..2015} and thermal conduction \citep{Orlando..et..al..2005, Bruggen..et..al..2016}. While these numerical works all contributed to our understanding of the cloud-wind interaction, almost all studies focusing on the problem of cloud acceleration and destruction studied either spherical or elliptical clouds.

Using extremely high-resolution simulations run with the GPU-based code \textit{Cholla}, \cite{Schneider..et..al..2015} explored the effects of spatial density distribution on the cloud-shock problem, following the evolution of both spherical clouds and of clouds with a lognormal density distribution set by turbulence. That study demonstrated that turbulent clouds are destroyed more quickly than spherical clouds of the same mass and mean density. In a followup study, \cite{Schneider..et..al..2017} ran a series of radiative simulations that explored the interaction between a hot ($T$ = $10^6\,\rm K$ ) supernova-driven wind and a cool turbulent cloud representing entrained interstellar material.

Due to the extremely high spatial resolution afforded by $Cholla$ and the realistic way that the turbulent clouds mix, these simulations are ideal for studying the phase structure of gas in outflows. After the initial shock, a wide distribution of densities and temperatures is created as gas in the cloud mixes with gas in the wind. This paper focuses on an analysis of those simulations at a characteristic time representative of an average state for a multiphase galactic wind. In particular, we analyze one of these cloud-wind simulations by creating mock absorption lines that can be compared to the plethora of observational data, thereby linking theoretical studies of cloud-wind interactions with observational studies of multiphase galactic outflows.

Section \ref{Cloud-Wind Simulation} describes the simulation of the interaction between a cool turbulent cloud and the hot wind. In Section \ref{Results} the physical properties of gas in the simulation are explored. Mock absorption lines are created for a variety of ionic species, and we compare different methods for deriving values of the optical depth and covering fraction. In Section \ref{Discussion}, we compare our results with previous observational and theoretical work. Finally, Section \ref{Conclusion} briefly summarizes the paper.

%This becomes a reason why most galactic outflow studies assume that the gas in the outflows is photoionized. Photoionized models do not reproduced the observed O VI, but adequately match the low ionization gas.  Our idealized simulation covers a large volume which allows us to keep track of the gas throughout its multiple phases. 

%\ES{I've updated the style file to AASTex version 6.2. The author guide to this LateX style guide is here: https://journals.aas.org/aastex-v6-2-author-guide/}

%\ES{Instead of using subject keywords, AAS has moved to a new UAT system (see https://journals.aas.org/aas-journals-uat/). We will select relevant concepts when we submit the paper. You should still update the keywords for the ArXiv post, I guess?}

\section{The Cloud-Wind Simulation} \label{Cloud-Wind Simulation}

%

%Explain cloud wind sim., box size (High res. GPU); throughout the section refer to fig. one. 

%In comparison to previous numerical work $Cholla$ allows us to study the different phases of gas over a larger region. By using $Cholla$ we can create an idealized simulation that captures the interaction between a hot supernovae driven wind $(n \approx 10^{-2.5} cm^{-3}, T \approx 10^6 K)$ a cool dense cloud $(n \approx 10^{2} cm^{-3}, T \approx 10^4 K)$ of interstellar media. 

In order to better understand the interaction between the hot and cool phases in outflows, \cite{Schneider..et..al..2017} ran a high resolution simulation of a hot
flow (with physical parameters modeling a supernova-driven wind) interacting with a cool, dense interstellar medium cloud. Below, we briefly describe the details of the simulation - a more extensive description can be found in the original work. 

The simulation was run using $Cholla$, a GPU-based hydrodynamics code that allows for efficient computation \citep{Schneider..et..al..2015}. The $Cholla$ code uses an HLLC Riemann solver and PPM reconstruction, and for this simulation optically-thin radiative cooling was implemented using pre-computed Cloudy tables \citep{Ferland..et..al..2013}. Cooling rates were calculated assuming solar metallicity for all the gas and cooling was allowed down to a temperature of $T = 10\,\rm K$. Allowing gas to cool to temperatures below $T = 10^4\,\rm K$ allowed for greater cloud compression, and substantially altered the density distribution relative to previous studies. Heating from a metagalactic UV background was also included \citep{Haardt2012}, although this does not make a practical difference to the thermodynamics for the regime studied. The simulation volume consists of a box with physical dimensions $160\times40\times40 \rm pc$, with $2048\times512\times512$ cells, yielding a resolution of $\Delta x = $ 0.07825 $\rm pc/cell$. The numerical efficiency of \textit{Cholla} 
enables this high resolution to be maintained across the entire simulation volume, allowing  us to study mixing and cooling and the evolution of the gas through multiple phases and across a large region.

Properties for the background hot wind were set based on the
\citet{Chevalier..and..Clegg..1985} spherical adiabatic solution, with an energy input rate $\dot{E}=10^{42} \ {\rm  erg\ s^{-1}}$, mass input rate $\dot{M}=2 M_\odot\ {\rm yr}^{-1}$, and driving region size $R_*=300\ {\rm pc}$, as motivated by the \cite{Strickland..et..al..2009} fit to Chandra X-ray observations of M82. The wind properties were set using the model solution at $r = 1$ $\rm kpc$, giving a number density $n_{\rm wind} = 5.26\times10^{-3}\,{\rm cm^{-3}}$, temperature $T$ = 3.7$\times 10^{6}\,\rm K$, and  velocity $v_{\rm wind} = 1.1962 \times 10^{3}\,\rm km s^{-1}$. Meanwhile, the cool component of the simulation is representative of interstellar material
recently entrained in the hot outflow. This cool material is set to be initially at rest and in thermal pressure equilibrium with respect to the background wind. The distribution of interior densities is set to a range of values as imposed by turbulence for Mach $\sim$ 5 on the scale of the cloud \citep{Schneider..et..al..2017}. The initial median density of the cloud is $\tilde{n} = 0.5,\\rm cm^{-3}$ and the median temperature is $T_{\rm med} = 3.98 \times 10^{4}\,\rm K$; the initial temperature distribution of the cloud is determined by the density. The cloud has an initial radius $R_{\rm cl} = 5 \rm pc$ , such that the column density of hydrogen through its center is $N_H = 1.02 \times 10^{20} {\rm cm^{-2}}$, and the total cloud mass is 8.61 $M_{\odot}$. The median density and inhomogeneous structure of the cloud influence its overall evolution, including the efficiency of cooling and the time it takes to be destroyed. The cloud has a low enough density that self-gravity can be ignored. 

To interpret and compare the results to previous models of cloud-wind interactions, we describe evolution
in terms of the cloud-crushing time, $t_{\rm cc}$ \citep{Klein..et..al..1994}. When the simulation starts, the wind drives a shock through the cloud, which causes the cool material to heat up and accelerate in the wind direction. The cloud-crushing time is an estimate of the time 
for the initial shock to propagate through the cloud and maximally compress it. This timescale can be calculated using the cloud-wind density ratio $\chi = \frac{n_{\rm cloud}}{n_{\rm wind}}$, the radius of the cloud $R_{cloud}$, and the wind velocity $v_{\rm wind}$:
\begin{equation}
    t_{\rm cc}= \frac{\chi^{\frac{1}{2}} R_{\rm cloud}}{v_{\rm wind}}.
\end{equation}
For the parameters of our simulation, $t_{\rm cc}= 39.8\,\rm Kyr$. The simulation was run for 1 Myr, equivalent to 25 cloud-crushing times. A real galactic wind would contain many clouds, at various stages of evolution; this simulation represents one small piece of that interaction. 

\begin{figure*}
\includegraphics[width=\linewidth]{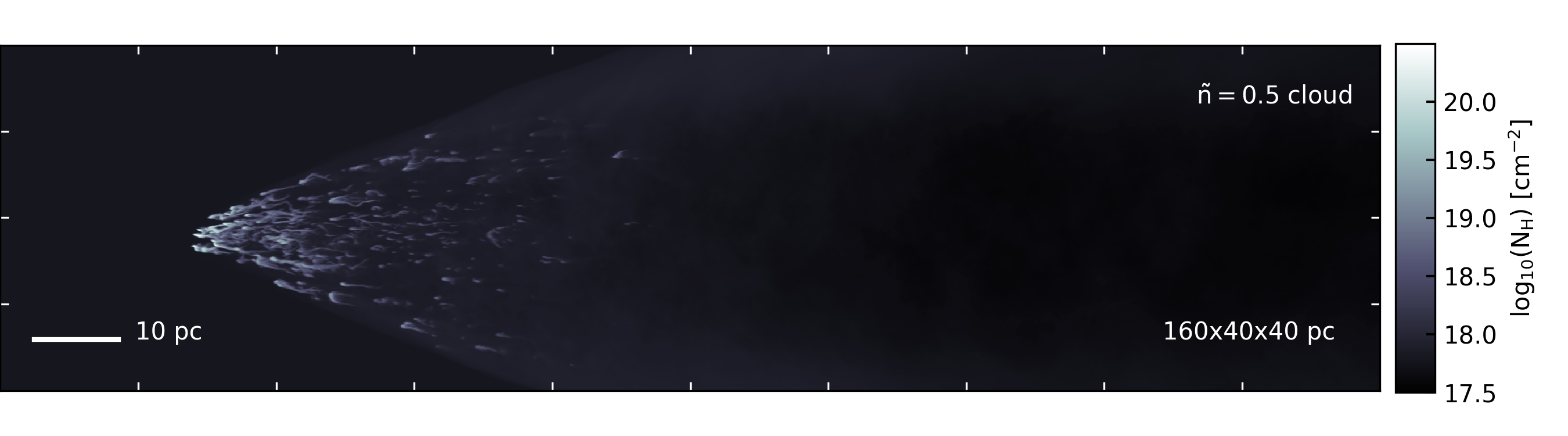}
\includegraphics[width=\linewidth]{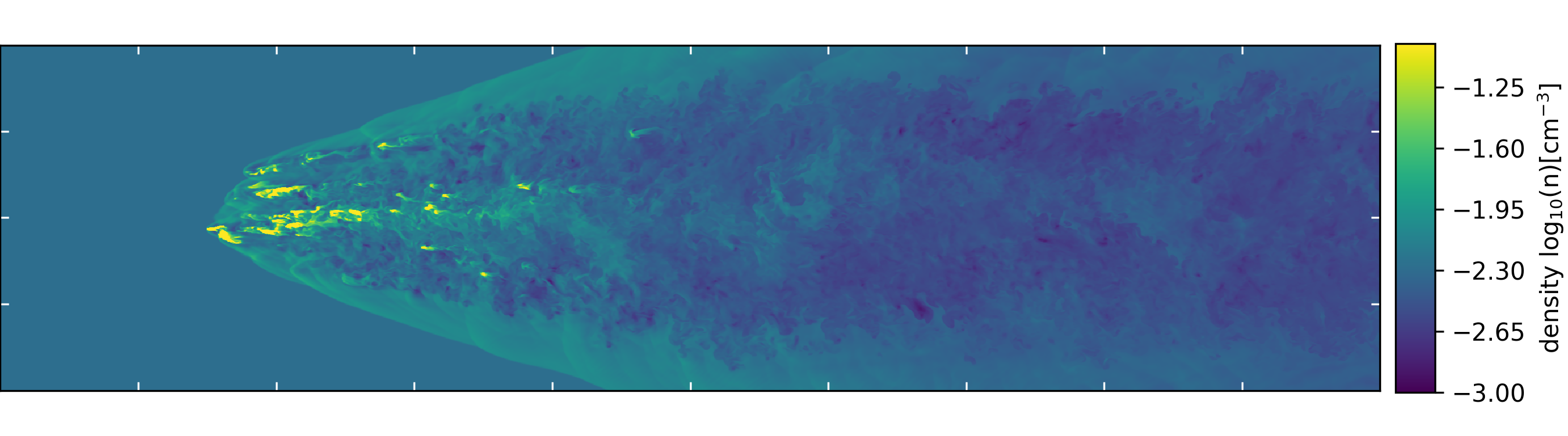}
\includegraphics[width=\linewidth]{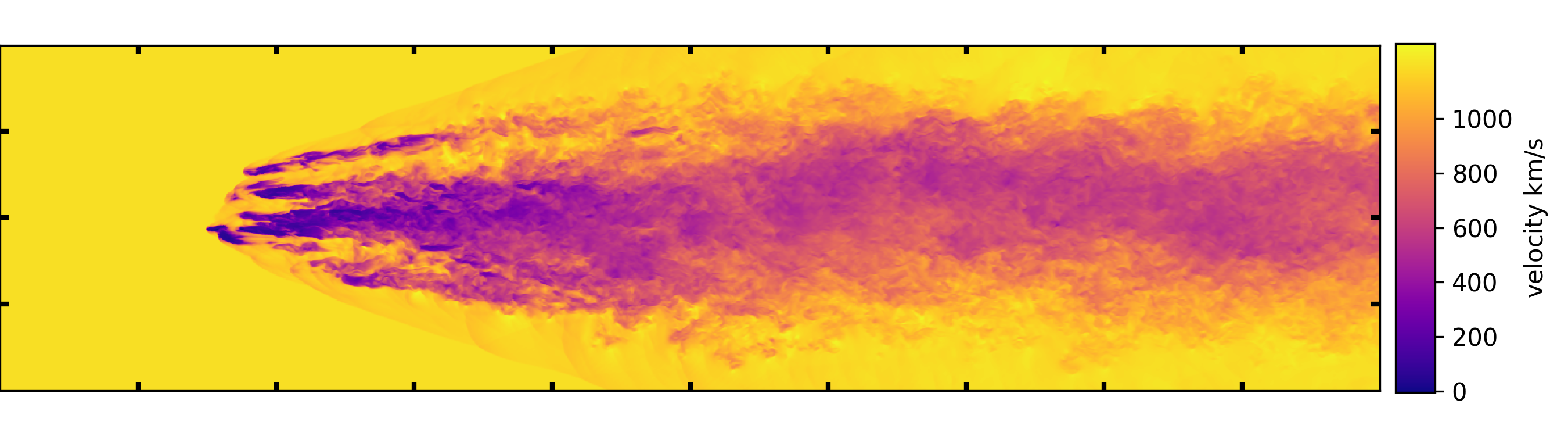}
\includegraphics[width=\linewidth]{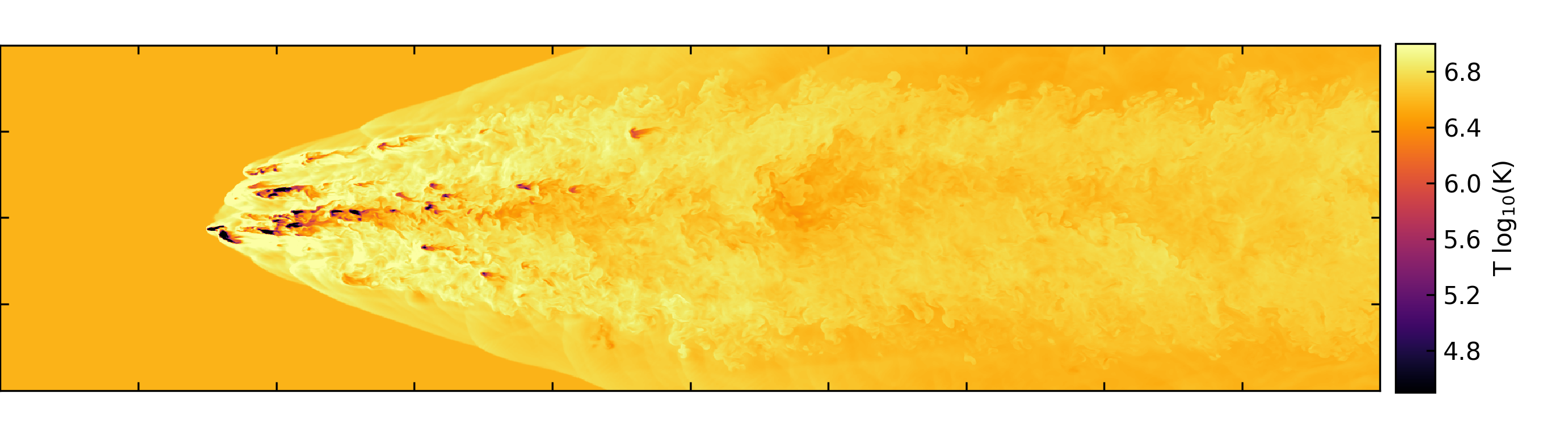}
\caption{Mixing between the hot and cool gas is captured at an intermediate state of $t_{\rm cc}=10$. Panels show the projected column density, as well as slices of density, velocity, and temperature  through the $x-z$ midplane.}
\label{fig:slices}
\end{figure*}

\section{Results}\label{Results}
We begin 
%our results 
with an overview of the physical properties of the gas throughout the simulation. We then discuss how the gas is split into different temperature ranges to represent different ions, as are often observed in real multiphase winds. Next, 
%a description of 
we describe 
how synthetic absorption lines 
were computed 
%is presented, 
%as well as the  
and present 
profiles for each ion. In calculating the absorption lines, we apply techniques used in observations that use doublets to obtain optical depth and covering fraction. We 
%conclude our results with a comparison of the 
then compare distributions of optical depth obtained via direct computation from the simulation, vs. those calculated using the empirically derived values.

\subsection{Mass, Velocity, and Temperature Distribution}

%\textbf{Figure \ref{fig:slices} allows us to see how the temperature, density and velocity are distributed throughout the simulation. In order to better understand this multiphase wind phenomenon we analyze the physical properties of the material.}

%\textbf{First we begin by studying how the gas is distributed throughout the simulation. We analyze the amount of mass and space that the gas occupies}.

%\textbf{From our figure\ref{fig:slices} we can see that the gas in our simulation has a range of temperatures,velocities and densities. We can see that the low temperature gas has a high density and  low velocity.} 

%\textbf{This leads us to separate the material in the simulation into different temperatures to represent different species. This is done by using collisionally ionized gas temperatures from \cite{2017ARA&A..55..389T}. Table \ref{table: ion_table} shows the six ionic species used in this study.}

%\textbf{Next we study the distribution of mass for three different species that vary in temperature. Studying the distribution for each particular ion allow us to see how much each species contributes to the total mass in the simulation.}

%\textbf{Furthermore, we study the column density of our three species along a particular line of sight. To construct the column density projections for each ion we assume solar metallicity throughout.From the column density we can see how the gas in distributed throughout the simulation. }

%EES

The cloud in this study reaches maximum compression at $t_{\rm cc}=2$ and completely disintegrates by $t_{\rm cc}=20$. We use $10 t_{\rm cc}$ for our analysis, as it represents an average state that adequately captures the mixing between the cool and hot gas. Figure \ref{fig:slices} shows a simulation snapshot at 10 $t_\mathrm{cc}$. The top panel shows column density projected on the $x-z$ plane. The lower panels show slices of the density, velocity and temperature through the center of the domain. 

The cloud material accelerates as momentum is transferred from the hot wind, and enters the outflowing wind with a range of velocities, as can be observed in the velocity slice. The low density material within the original cloud accelerates much faster than the high density material, and we see that velocities are anticorrelated with density. The density slice shows that most of the volume is occupied by low density gas, while the high density gas is clumped together in small knots, generally seeded by the highest-density regions of the original turbulent cloud. The temperature slice shows that most of the volume is filled with high temperature gas at approximately $ T=10^{6.5}\rm K$, the temperature of the shocked hot wind. By contrast, the highest density regions remain at low temperature, driven by efficient radiative cooling of the highest density cloud gas. By volume,  most of the gas in the simulation has a low density, high velocity, and high temperature, but it should be noted that this is partly driven by our selection of a small initial cloud.

\begin{figure}[t!]
\centering 
\includegraphics[width=\linewidth]{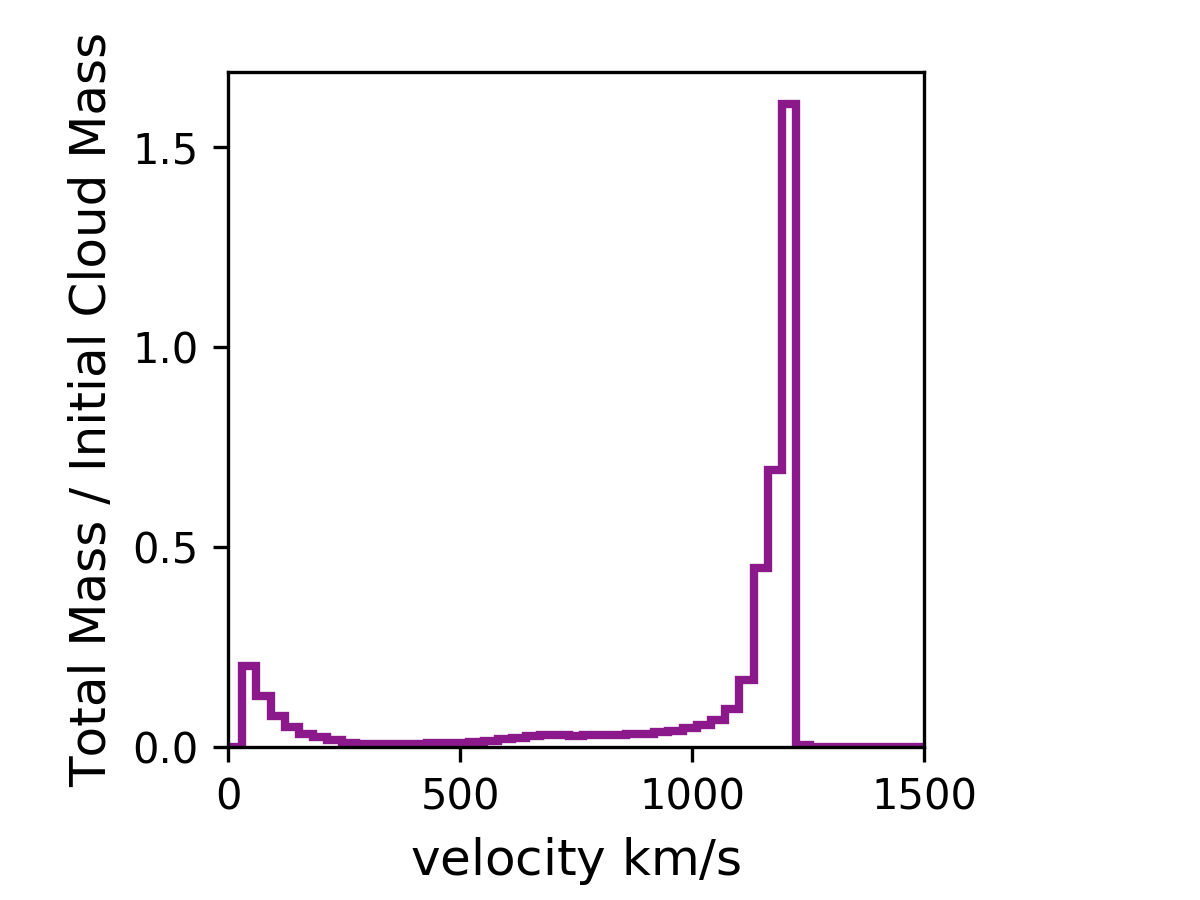}
\includegraphics[width=\linewidth]{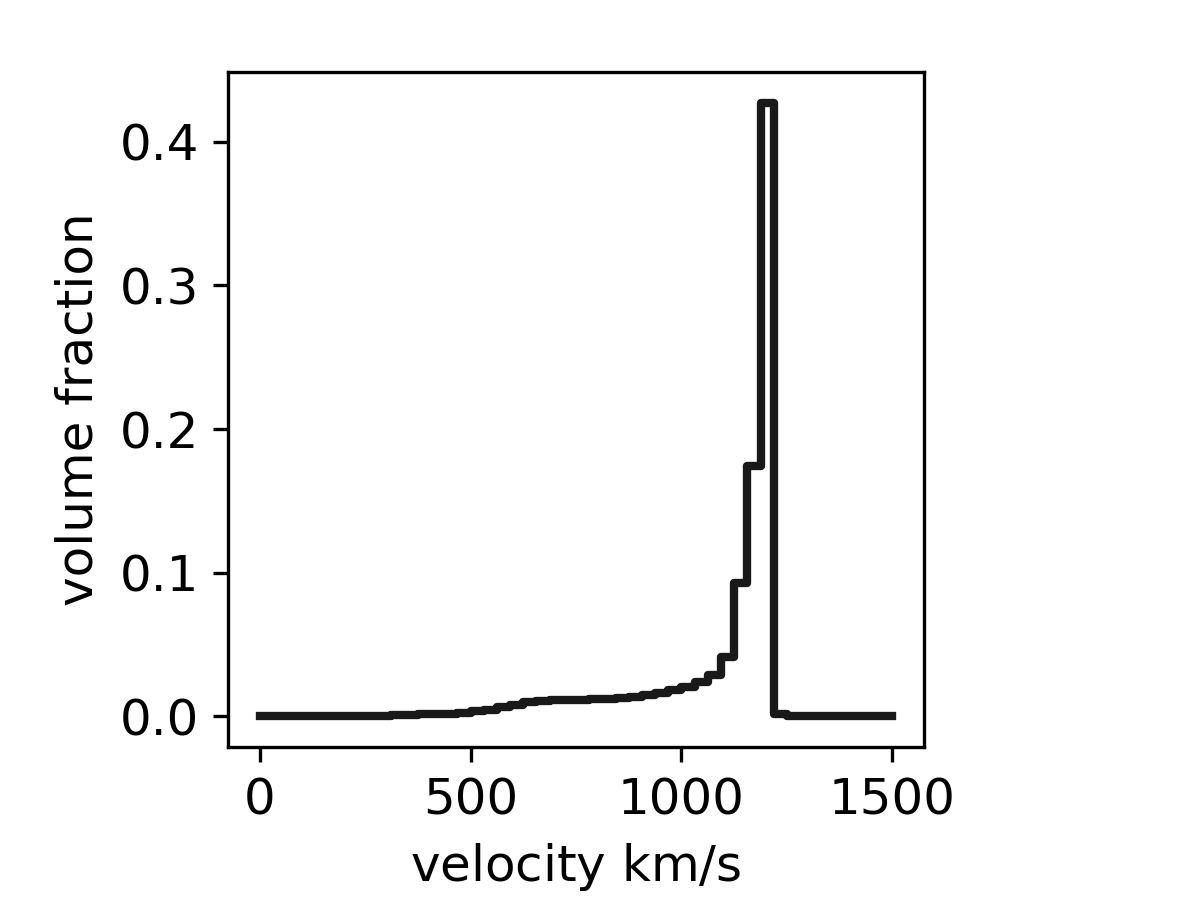}
\caption{1D mass-weighted (top panel) and volume-weighted (bottom panel) velocity histograms. The hot gas travels at high velocities ranging from $1,000$ to $1,250 \rm km \ s^{-1}$  and comprises the majority  of the mass and almost all of the volume. The cool cloud material occupies a negligible amount of space, but has a significant mass contribution.}
\label{fig:MVHistogram}
\end{figure}

The velocity distributions of the gas in the simulation can be quantified via one-dimensional histograms. The distribution of all the gas in the simulation is shown in Figure~\ref{fig:MVHistogram}, in mass-weighted (top) and volume-weighted (bottom) 1D velocity histograms. The upper panel shows clear peaks that differentiate the cool and hot gas phases. By mass, about 90 \% of the gas in the simulation is part of the hot wind (defined by an $T > 2 \times 10^{4} \rm K$), as expected given the small initial cloud mass. At this particular cloud-crushing time of 10 $t_{\rm cc}$, we find that the total mass of gas at $T < 2 \times 10^4\,\rm K$ is approximately 43\% of the initial cloud mass. The hot wind moves at a high velocity, $\sim$ 1,000 $\rm km \ s^{-1}$ to 1,250 $\rm km \ s^{-1}$. The cooler gas, created by mixing with the cloud, has velocities in the range 30-200 $\rm km \ s^{-1}$. The ratio of mass in low to high velocity components reflects our choice of cloud mass to the mass of hot gas that fits within the domain. The bottom panel in Figure~\ref{fig:MVHistogram} shows the relative contribution of the hot wind and cool cloud to the total volume in the simulation. We can see from this Figure that 40\% of the gas is unaffected wind material moving at the initial velocity of $1,200 \rm km \ s^{-1}$.

As a simple way to map the gas in our simulation to different absorption lines, we follow Figure 6 in  \cite{Tumlinson..2017}, in which temperature ranges for each ion are given based on the full width at half maximum (FWHM) surrounding the temperature associated with  the peak abundance of that ionic species from collisional ionization. Table~\ref{table: ion_table} shows the range of temperatures adopted for each species in this study.
%In order to probe different ionization states of the gas, we use peak values of collisionally ionized gas to assign ionization states to gas in different temperature ranges. Following Figure 6 in  \cite{Tumlinson..2017}, temperature ranges for each ion are selected according to the full width at half maximum (FWHM) surrounding the peak temperature for collisional ionization of that ionic species, as shown in Table~\ref{table: ion_table}. 
Assigning ionization states in this way allows us to study the contribution of each ionic species to the total mass and volume throughout the simulation.
\begin{table}
\caption{Ionic Species Used in Absorption Line Analysis} \label{table: ion_table}
\begin{tabular}{|c | c | c | c | c | c | c|}
\hline
Ion & $ \rm log_{10}$($T$) & $\lambda_{0,w}$ & $\lambda_{0,s}$ & $f_w$ & $f_s$ & $\frac{n}{n_{H}}$\\
\hline
 Si II  & 3.8 - 4.3 & 1193 & 1260 & 0.5820 & 1.1800 & $3.47e^{-5}$\\
 C II   & 3.9 - 4.9 &  904 & 904 &  0.1680 & 0.3360 & $2.45e^{-4}$\\
 Si IV  & 4.6 - 5.2 & 1403 & 1394 &  0.2600 & 0.5240 & $3.47e^{-5}$\\
 C IV   & 4.9 - 5.3 & 1551 & 1548 & 0.0952 & 0.1900 & $2.45e^{-4}$\\
 N V    & 5.1 - 5.5 & 1243  & 1239 &  0.0777 & 0.1560 & $8.51e^{-5}$\\
 O VI   & 5.3 - 5.7 & 1038 & 1032 &  0.0658 & 0.1325 & $4.90e^{-4}$\\
\hline
\end{tabular}
\\
Col. 1: Ion Species
\\
Col. 2: Gas temperature range [K], calculated using the full-width half max range for collisionally ionized gas (see e.g. Table 6 of \cite{Tumlinson..2017})
\\
Col. 3: Rest-wavelength for weak absorption lines.
\\
Col. 4: Rest-wavelength for strong absorption lines.
\\
Col. 5: Oscillator strength for weak absorption lines.
\\
Col. 6: Oscillator strength for strong absorption lines.
\\
Col. 7: Solar abundance (relative to hydrogen) for each ionic species.
\end{table}

\begin{figure*}[!htb]
\centering
$
\begin{array}{ccc}
\includegraphics[width=0.32\linewidth]{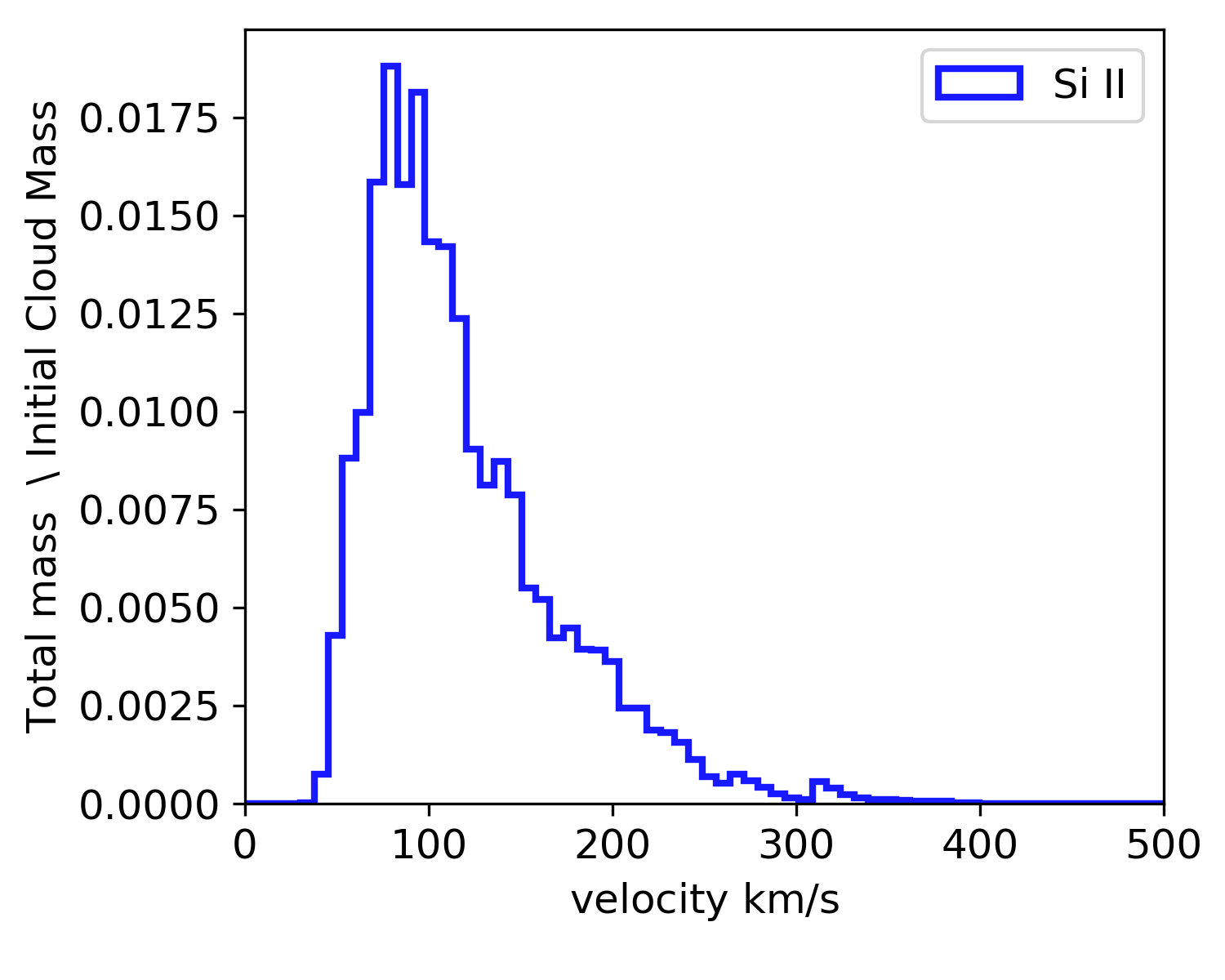}
\includegraphics[width=0.32\linewidth]{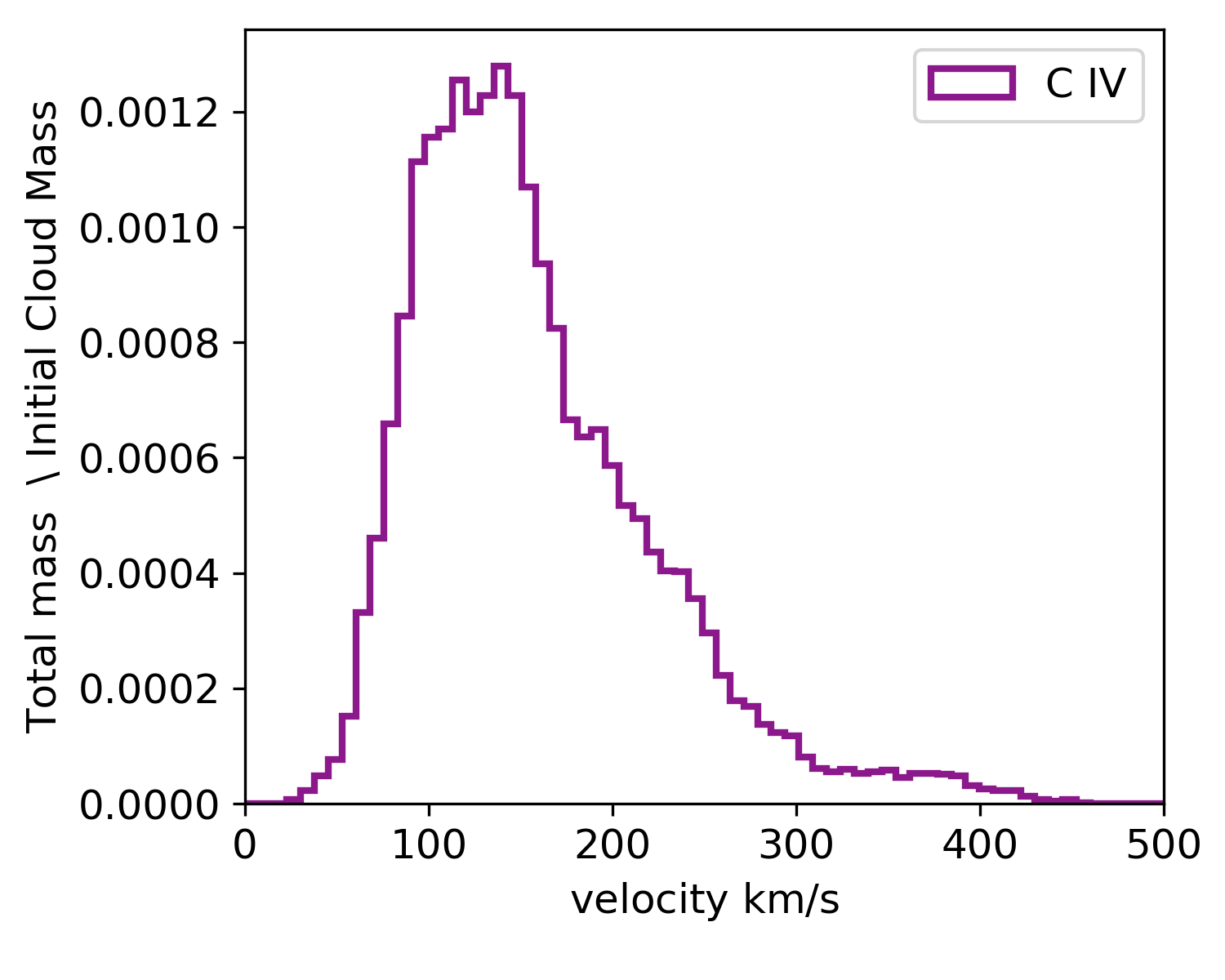}
\includegraphics[width=0.32\linewidth]{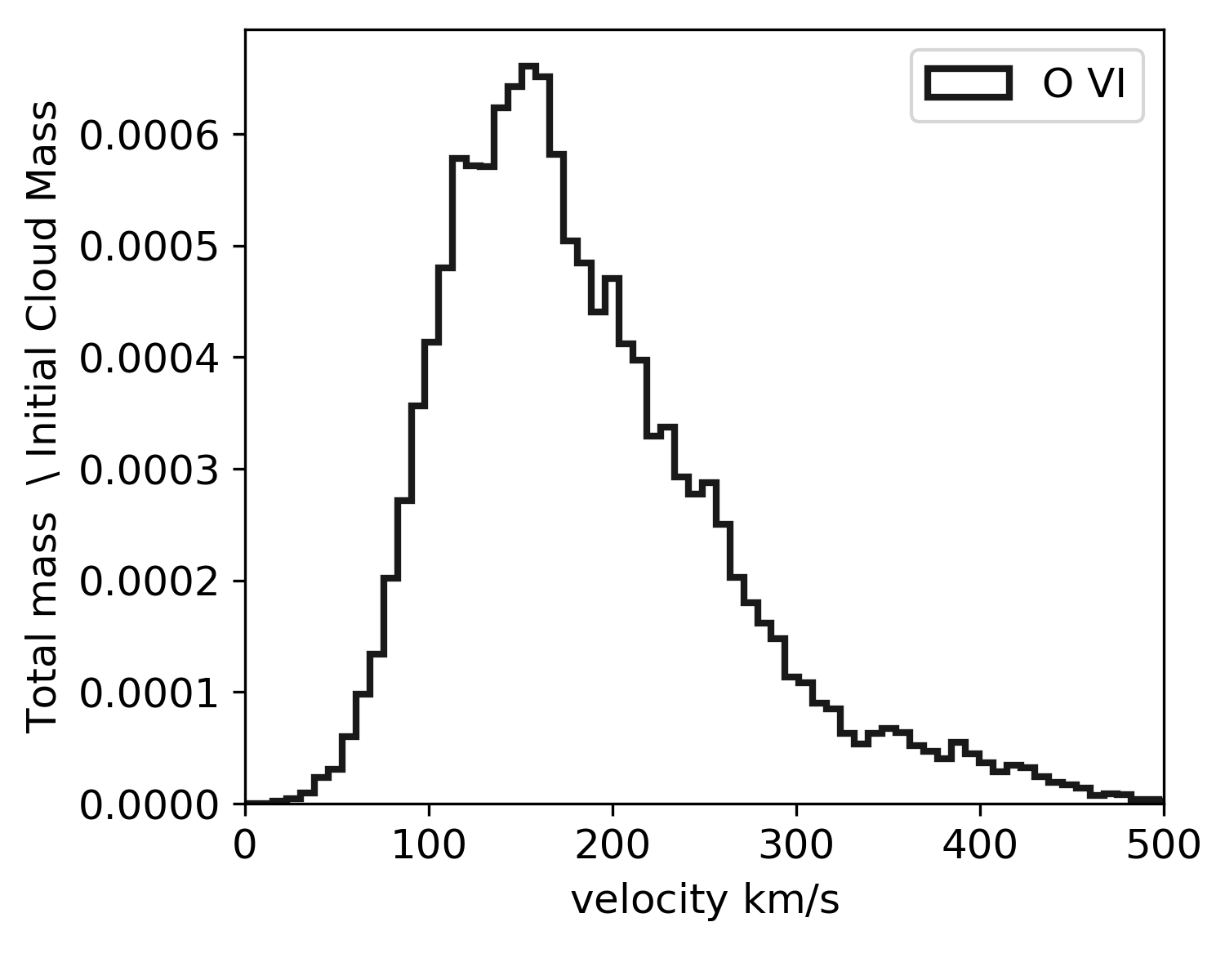}
\end{array}
$
\caption{The distribution of mass with velocity for gas traced by a low, intermediate, and high temperature ion. The mass decreases for increasing temperature/ionization state, while the range of velocities increases.
%As temperature of these ions increases their mass contribution decreases 
}
\label{fig:1Dhistograms}
\end{figure*}

In Figure \ref{fig:1Dhistograms}, we show mass-weighted 1D velocity histograms for a representative low, medium, and high ion constructed using these temperature ranges. In general, we see similar behavior between low, intermediate, and high ions, and find that Si II, C IV, and O VI are representative of other ions in those ranges. These histograms show that the higher ionization species contribute less to the total mass in the simulation. Furthermore, we see that the dense gas traced by the low ions has slower average velocities.

\begin{figure*}[!htb]
\centering
\includegraphics[width=0.29\linewidth]{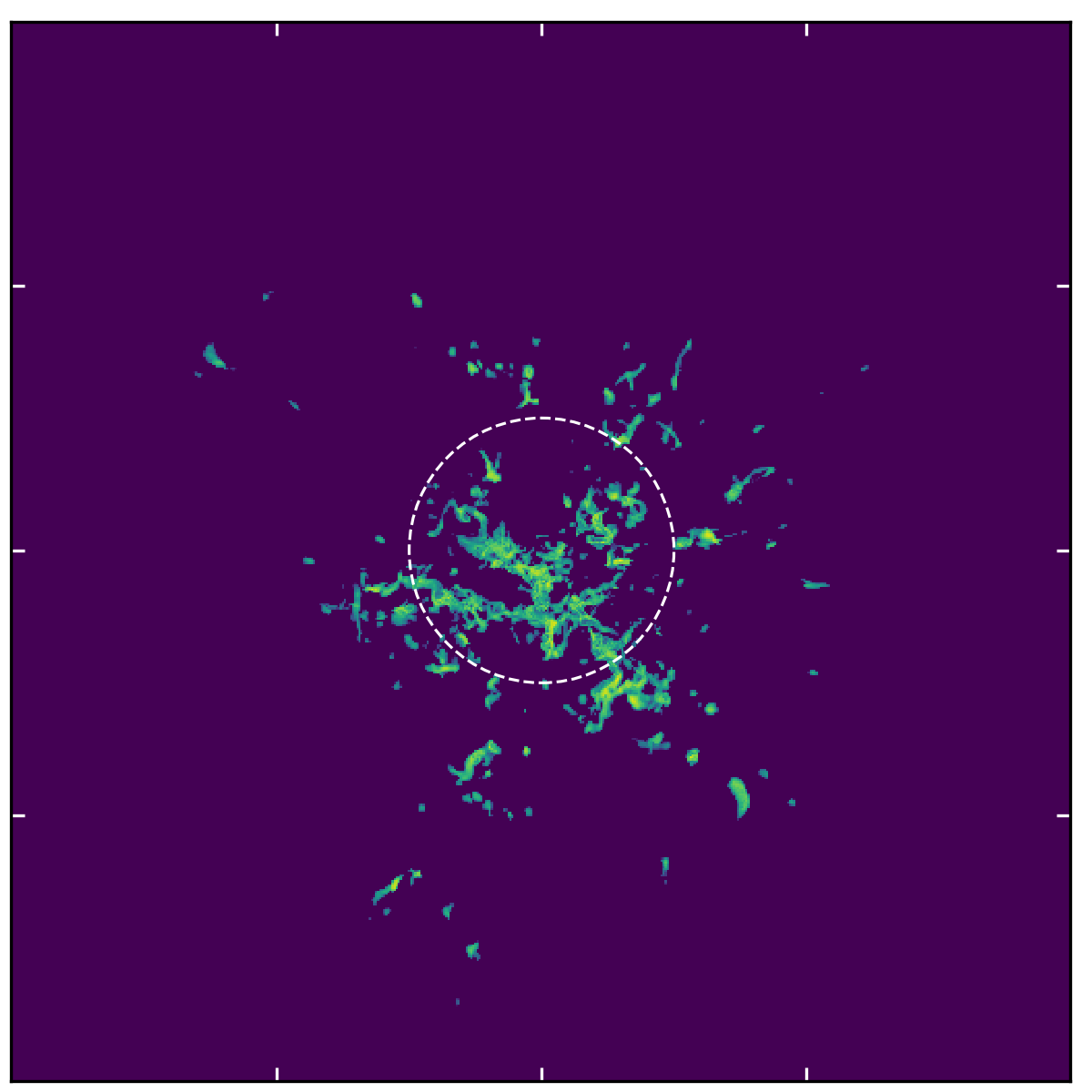}
\includegraphics[width=0.29\linewidth]{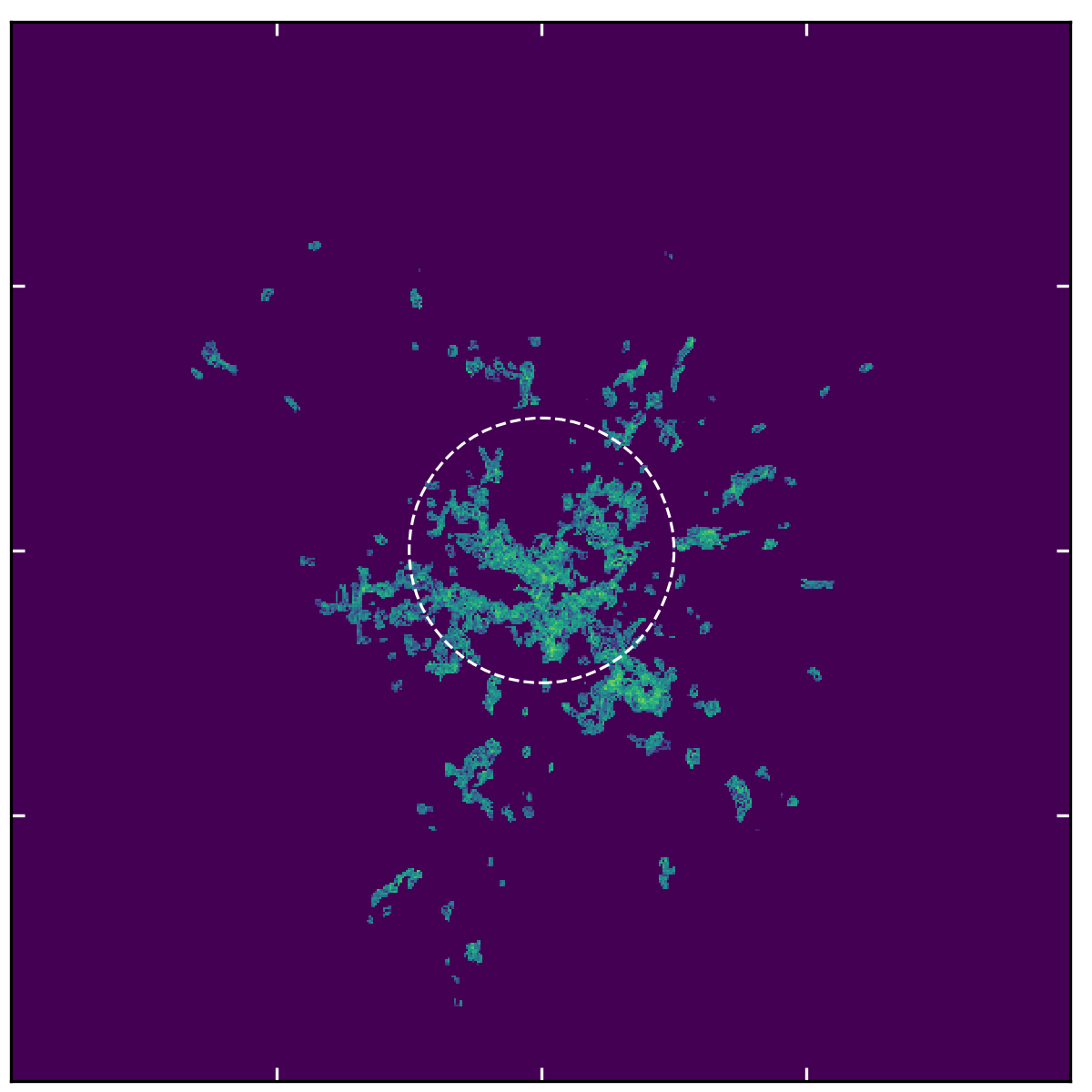}
\includegraphics[width=0.29\linewidth]{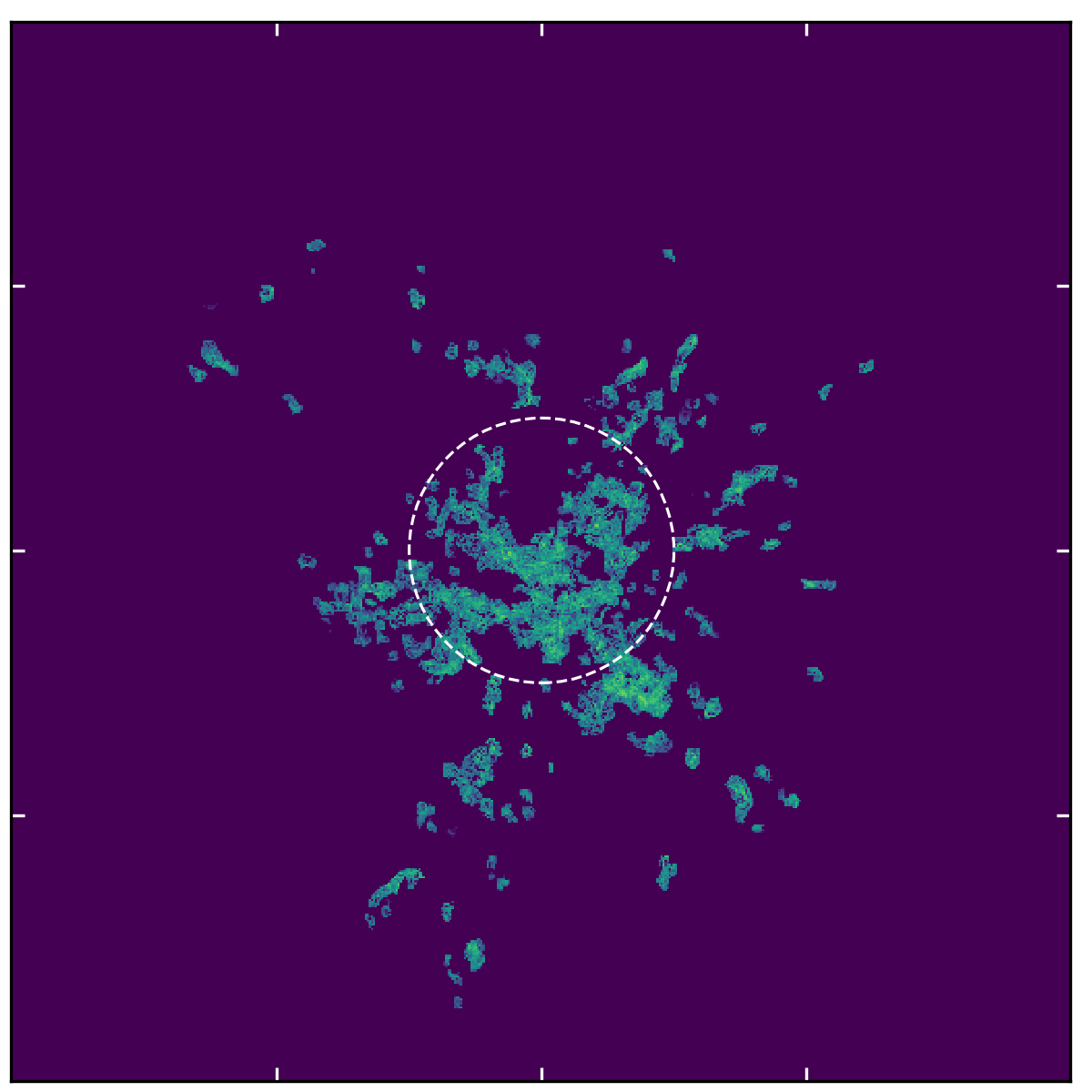}
\includegraphics[height=6.2cm]{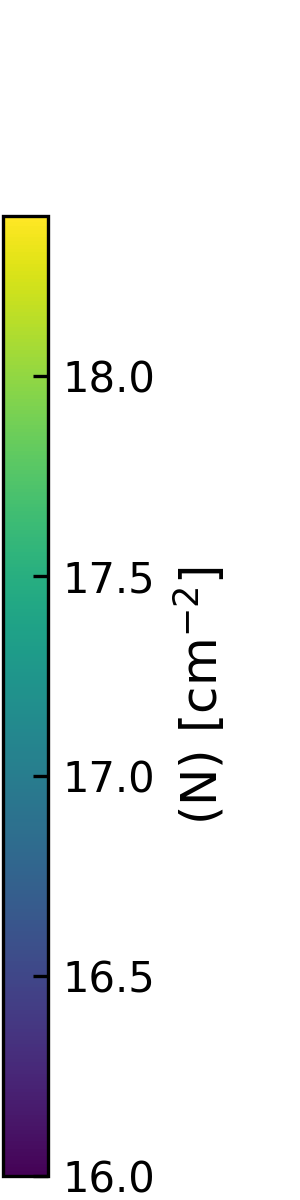}
\caption{Column density maps of representative low, intermediate, and high temperature ions. %for a $\tilde{n} = 0.5$ turbulent cloud at 10 $t_{cc}$. 
The white dashed circle represents the original extent of the cloud in the simulation. The higher ions 
%tend to 
have slightly lower column densities and are slightly more spatially extended.}
\label{fig:projections}
\end{figure*}

Adopting the abundances in Table~\ref{table: ion_table}, Figure \ref{fig:projections} shows maps  of column density for the three representative low, medium, and high ions. In general, O VI tends to be more spatially extended than the low ionization ions. Still, the three ions trace each other fairly well, indicating that the gas giving rise to them is generally spatially related. This reinforces the idea that the intermediate and higher ionization species are primarily being created as a result of the mixing between the cool cloud and the hot wind. 

%%%%%%%%%%%%%%%%%%%%%%%%%%%%%%%%%%%%%%%%%%%%%%%%%%%%%%%%%%%%%%%%%%%%%%%%%%%%%%%%%%%%%%%%%

\subsection{Absorption Line Profiles}

%\textbf{Previous work has studied absorption line profiles from photoionized gas. In this work we study synthetic absorption lines from collisionally ionized gas temperatures.}

%\textbf{To create synthetic absorption lines we study the optical depth for doublets of the same ionic species. The optical depth for a given doublet can be related by $\tau_s(v)=2\tau_w(v)$. }

%\textbf{We use our computed $\tau(v)$ for both strong and weak lines and integrate over the entire size of the simulation box \ref{eq: normalized flux}.}

%\textbf{Figure \ref{fig: Flux} shows us the comparison between strong and weak line profiles. }

%Previous studies of outflows pivot around a warm $10^{4} K$ gas. Mainly, because this gas phase has strong absorption and emission lines in the rest-frame ultraviolet and optical \cite{2018MNRAS.474.1688C}. This becomes a reason why most galactic outflow studies assume that the gas in the outflows is photoionized. Photoionized models do not reproduced the observed O VI, but adequately match the low ionization gas.  Our idealized simulation covers a large volume which allows us to keep track of the gas throughout its multiple phases. 

We study absorption line profiles of six ions that probe different temperatures. As demonstrated by the temperature slice in Figure~\ref{fig:slices}, gas in the simulation has a range of temperatures from $10 - 10^{7} \rm K$ . We create mock absorption line profiles using the optical depth of each ionic species, which reflects both  total column densities in each map pixel (e.g. Figure~\ref{fig:projections}) as well as the line-of-sight velocity distribution. The optical depth is computed adopting the Sobolev approximation, which gives $\tau$ in terms of the density and radial velocity gradient as
\begin{equation}\label{eq: tau}
    d\tau(v;y,z)=\frac{\pi e^2}{\rm mc} f \lambda_0 n(v;x,y,z) \frac{\rm dx}{\rm dv}.
\end{equation}
Here, the line-of-sight velocity gradient is taken to be in the direction of the background wind ($x$, in the simulation). The oscillator strength ($\rm f$) and rest-wavelength ($\lambda_0$) of each of the six species investigated in this study can be found in Table \ref{table: ion_table}.  By integrating the number densities along the $x$-axis in bins of velocity, we can directly compute the optical depth along each line-of-sight, giving $n_y\times n_z = 262144$ independent sightlines through the simulation box. We use these to calculate the total flux decrements as a function of velocity for each species through the box; we shall also use the spatial distribution to assess covering factors.

\begin{figure*}[!htb]
\centering
\includegraphics[width=0.32\linewidth]{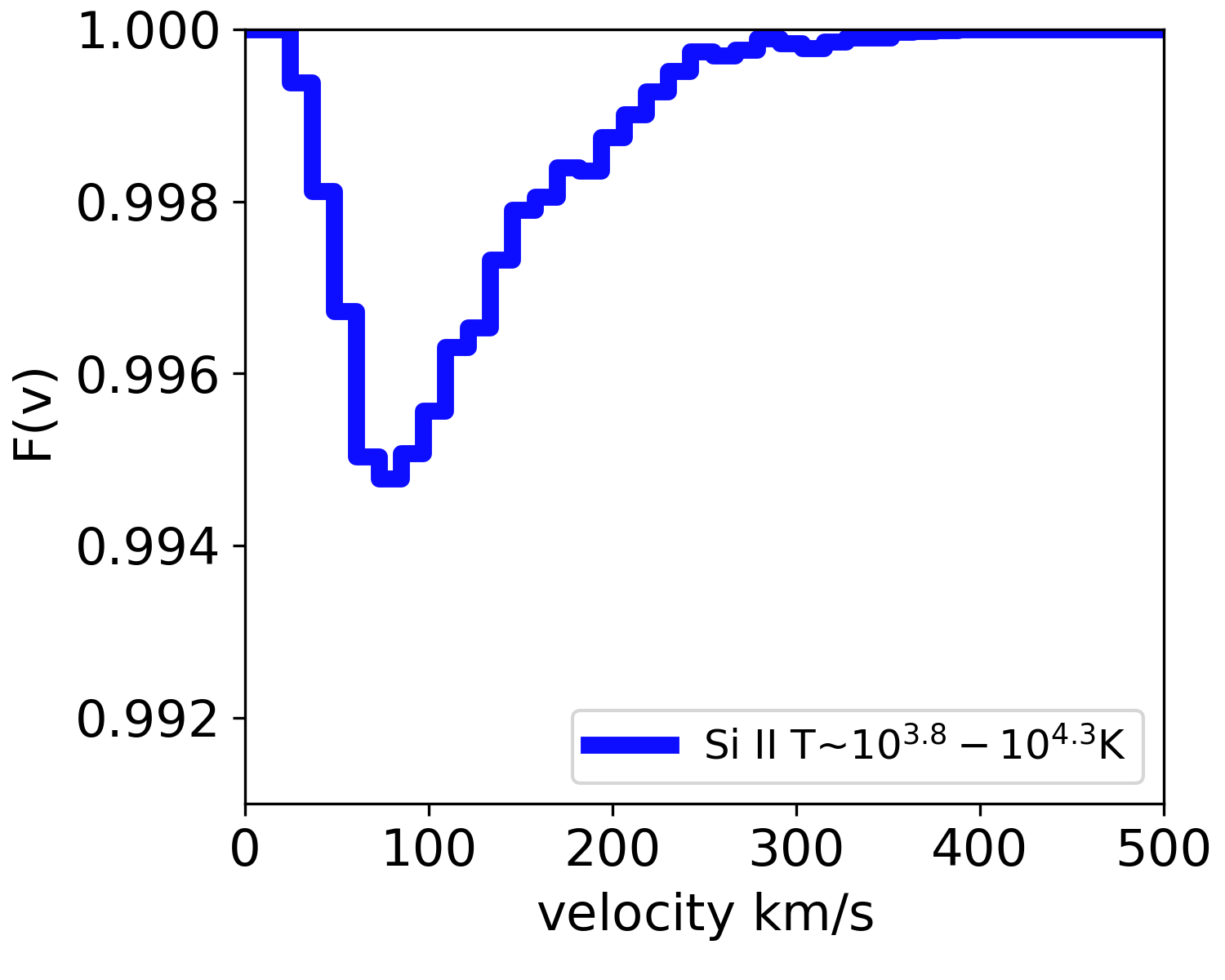}
\includegraphics[width=0.32\linewidth]{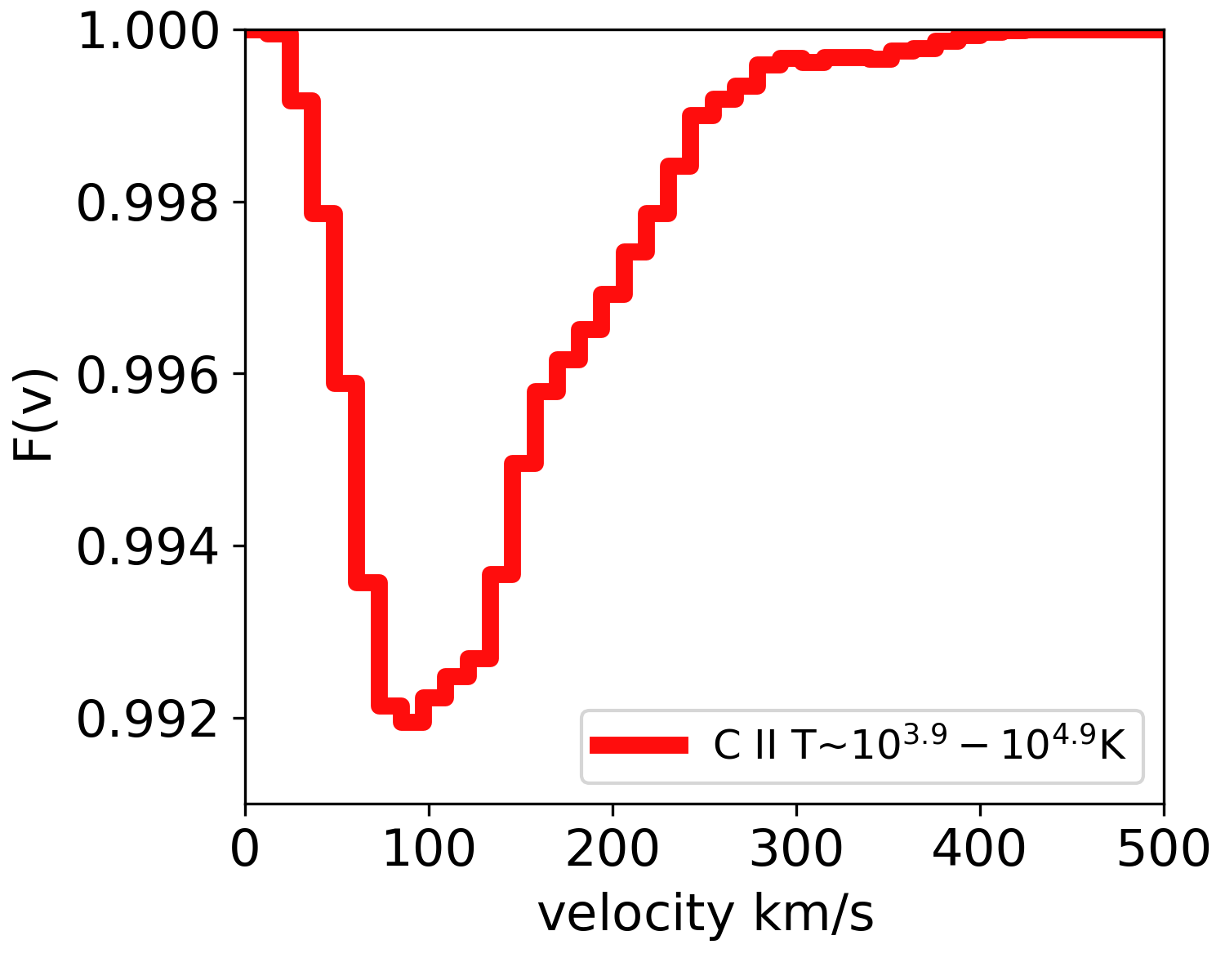}
\includegraphics[width=0.32\linewidth]{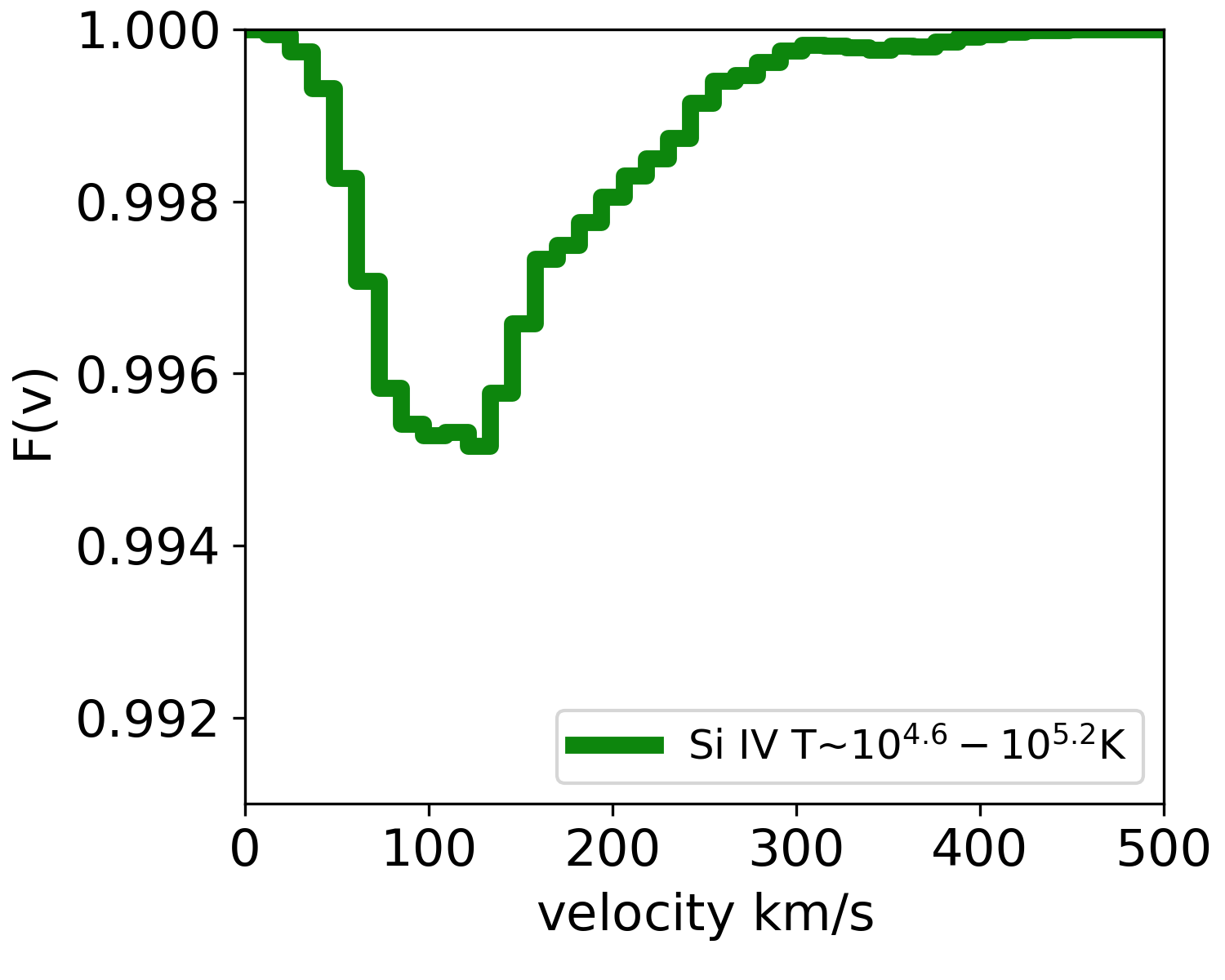}
\includegraphics[width=0.32\linewidth]{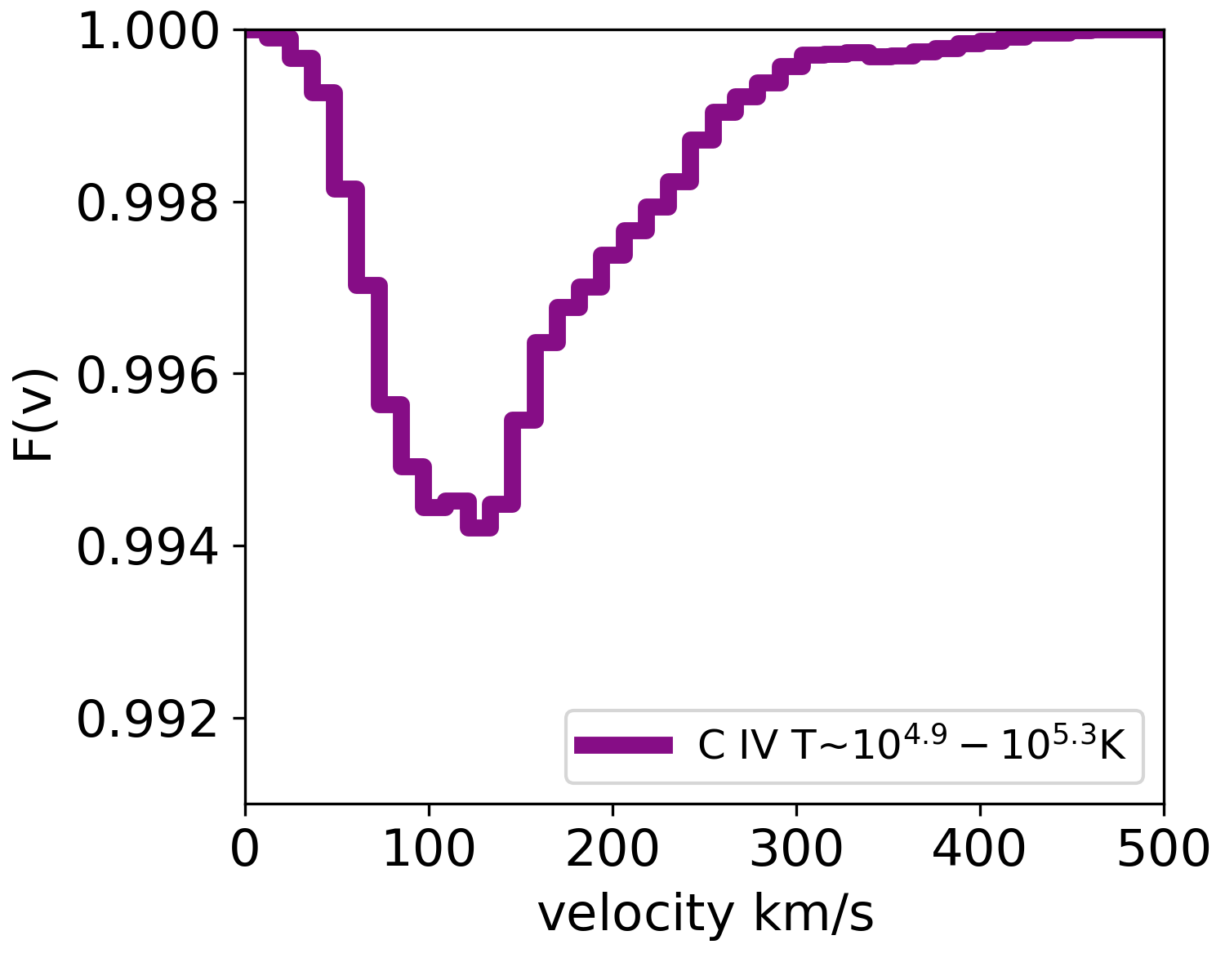}
\includegraphics[width=0.32\linewidth]{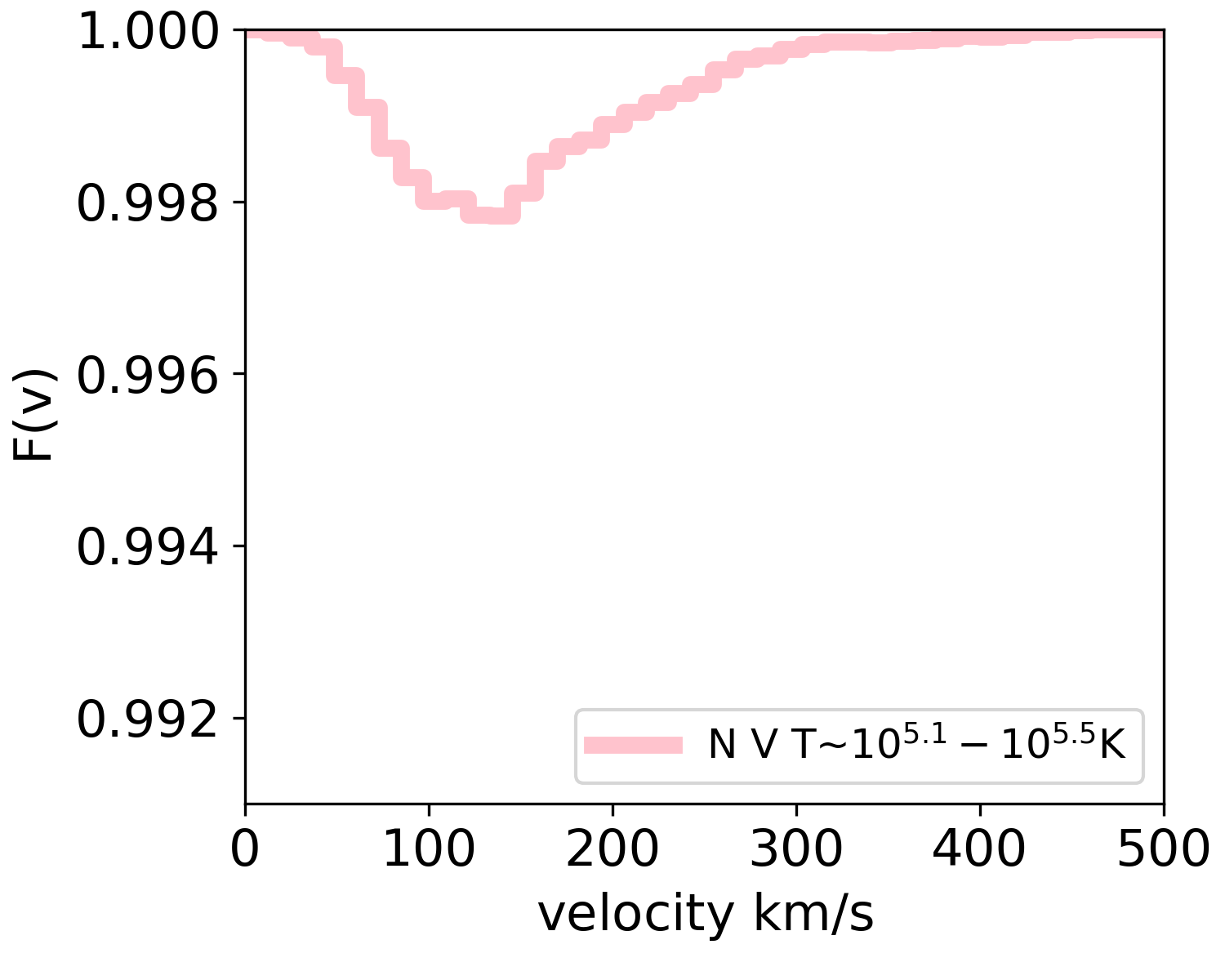}
\includegraphics[width=0.32\linewidth]{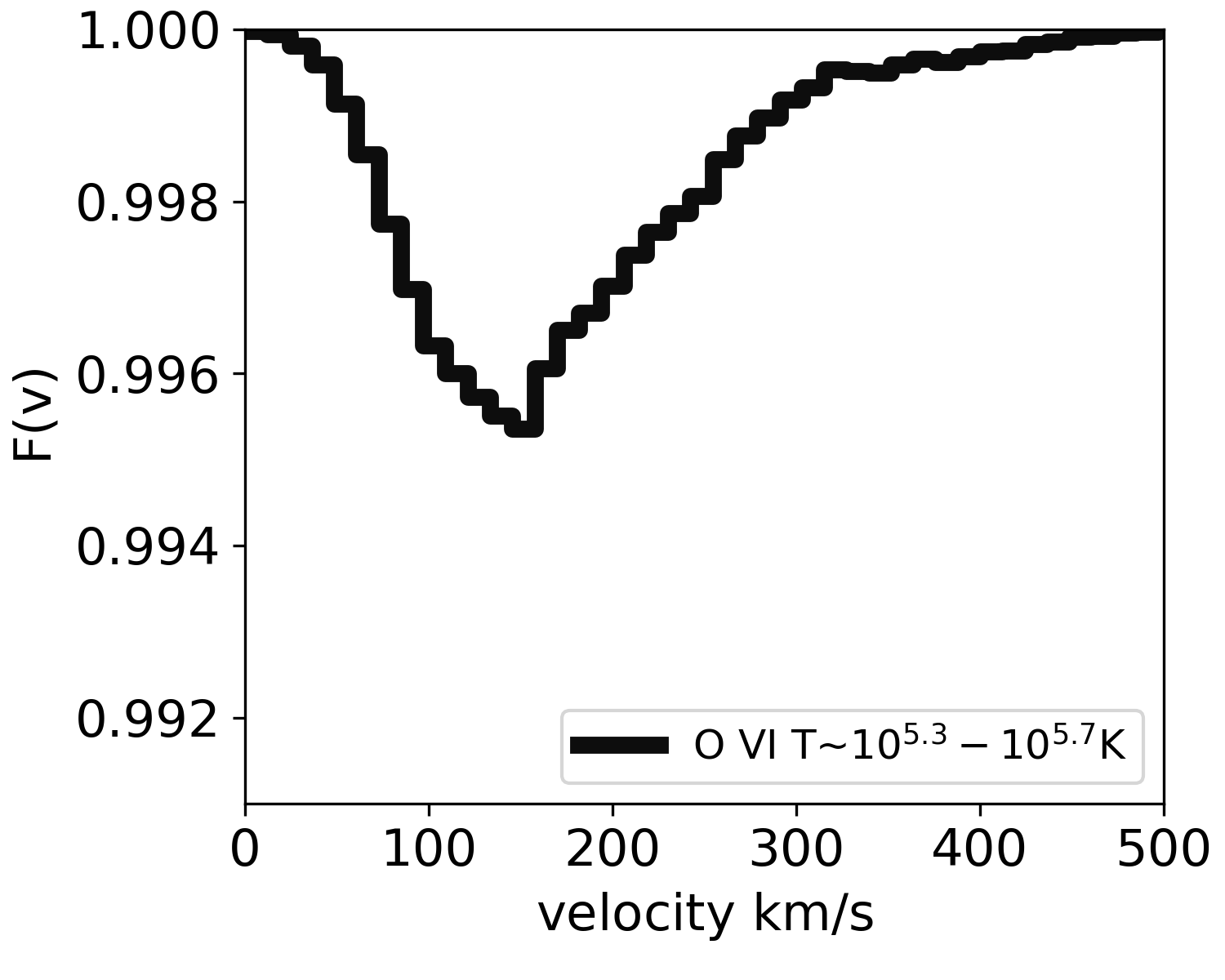}
\caption{Normalized synthetic absorption line profiles for commonly-observed low, intermediate, and high ions. Table \ref{table: velocity table} lists the variation of $v_{\rm min}$, $v_{\rm central}$, and $v_{90}$ for each of the lines shown here.}
\label{fig: Flux}
\end{figure*}

The normalized observed flux $F(v)$ is defined as
\begin{equation}\label{eq:Direct Flux}
    F(v)=\frac{\int dy dz e^{-\tau(v;y,z)} S}{\int dy dz S}= \langle e^{-\tau(v)} \rangle_{\rm yz},
\end{equation}
where $S$ is the background source brightness (a stellar continuum) and where the average is over the $y-z$ projection of the simulation box. In the limit of low optical depth,
\begin{equation}
F(v) \approx 1 - \langle \tau(v) \rangle_{yz};    
\end{equation}
from the definition of \autoref{eq: tau} this involves spatial integration over the whole box. The flux decrement is then  proportional to $M(v)$, the distribution of mass as a function of velocity traced by the ion.   

Figure \ref{fig: Flux} shows the normalized flux as a function of velocity for the six ions in Table~\ref{table: ion_table}. 
%In reality, the observed flux comes from stellar continuum of the background galaxy, while absorption is the result of intervening clouds. 
Typically, down-the-barrel studies do not spatially resolve individual regions of galaxies, and thus the absorption is a result of the integrated continuum and absorption across the entire central starburst region of a galaxy. For our normalized fluxes, the averaging is done over the entire box, so the size and mass of the cloud relative to the simulation box determines how much flux we observe. The larger the simulation box, the more unobstructed sight lines there are in the average, leading to shallower absorption lines. The bigger the clouds and the higher the column density, the more high $\tau$ sightlines enter the average, and the smaller the directly computed flux. 

Three commonly measured features of absorption lines in winds are their minimum, central, and 90th percentile velocities. For each of the six ionic species, Table \ref{table: velocity table} lists $v_{\rm min}$, which is the velocity value corresponding to the minimum flux in the absorption line. $v_{\rm central}$ is a similar measurement, but is computed by first integrating the flux profiles (see Figure \ref{fig: Flux}) from 0 to 1200 km/s and identifying the velocity at which we reach 50\% of the area under the curve. Differences between $v_\mathrm{min}$ and $v_\mathrm{central}$ give some idea as to the non-Gaussian skew of the absorption line - $v_\mathrm{central}$ is higher than $v_\mathrm{min}$ for all of the lines, reflecting the extended high velocity tails. $v_{90}$ is computed similarly to $v_\mathrm{central}$ with the exception that it identifies the velocity that corresponds to 90\% of the area under the curve, giving a rough measure of the maximum (observable) velocity of the gas in the wind. Table \ref{table: velocity table} gives these values for each of our selected ions based on the profiles presented in Figure~\ref{fig: Flux}.  From Table \ref{table: velocity table} we can see an increase in all three velocity statistics as we go from low ionization ions (Si II and C II) to high ionization ions (N V and O VI).

\begin{table}
\caption {Velocity Properties of Synthetic Lines} \label{table: velocity table}
\begin{tabular}{|l | c | c | c | c | c | c |}
\hline
Ions & Si II  & C II & Si IV & C IV & N V & O VI\\
\hline
$v_{min,s}$ &  85 & 97 & 133 & 133 & 145 & 158 \\
$v_{central,s}$ & 132 & 144 & 156 & 156 & 168 & 180 \\
$v_{90,s}$ &  228 & 240 & 264 & 264 & 276 & 300\\
\hline
\end{tabular}
\\
$v_{\rm min}$, velocity value corresponding to minimum flux value.
\\
$v_{\rm central}$, velocity value corresponding to 50\% of the integrated flux.
\\
$v_{90}$, velocity value that corresponding to 90\% of the integrated flux.
\end{table}

%It is concluded that all velocities except those from the intermediate ions increase as temperature increases.
%%%%%%%%%%%%%%%%%%%%%%%%%%%%%%%%%%%%%%%%%%%%%%%%%%%%%%%%%%%%%%%%%%%%%%%%%%%%%%%%%%%%%%%%%

\subsection{Empirical Covering Fraction and Optical Depth}\label{Optical Depth and Covering Fraction}

%\textbf{By applying the techniques used in observations we are able to create empirical distributions of optical depth and covering fraction for each of our ionic species.}

%\textbf{First we begin by using our normalized flux represented by equation to compute the flux for the weak and strong lines of each ion.}

%\textbf{Using the value from table \ref{table: ion_table} and our column density values we can compute the optical depth for each species by using equation \ref{eq: tau}.}

%\textbf{The resulting optical depth values that are obtained for each ionic species is then used to compute the flux.}

%\textbf{The flux allows us to compute the empirical covering fraction and empirical optical depth for each ion. These can be computed by using \ref{eq: empirical} and \ref{eq: covering fraction}}

The depth of an absorption line depends on both the optical depth $\tau$ and the covering factor $C_f$ of the absorbing gas. The covering factor accounts for the fraction of background light source which is obscured by the absorbing gas, while the optical depth measures the total amount of absorption in a given sight line as light passes through the cloud. For example, in Figure~\ref{fig: Flux}, the shape of the absorption lines is primarily determined by differences in $\tau$ as a function of velocity, while the overall normalization is set by the $y-z$ spatial extent of our cloud (see Figure \ref{fig:projections}). If lines are optically thin, the absorption profiles are essentially rescaled, inverted profiles $M(v)$, examples of which are shown in \autoref{fig:1Dhistograms}. When $\tau > 3$ we define the line profile as optically thick, at which point the depth of the observed line will be set exclusively by $C_f$. 

\begin{figure}
\includegraphics[width=\linewidth]{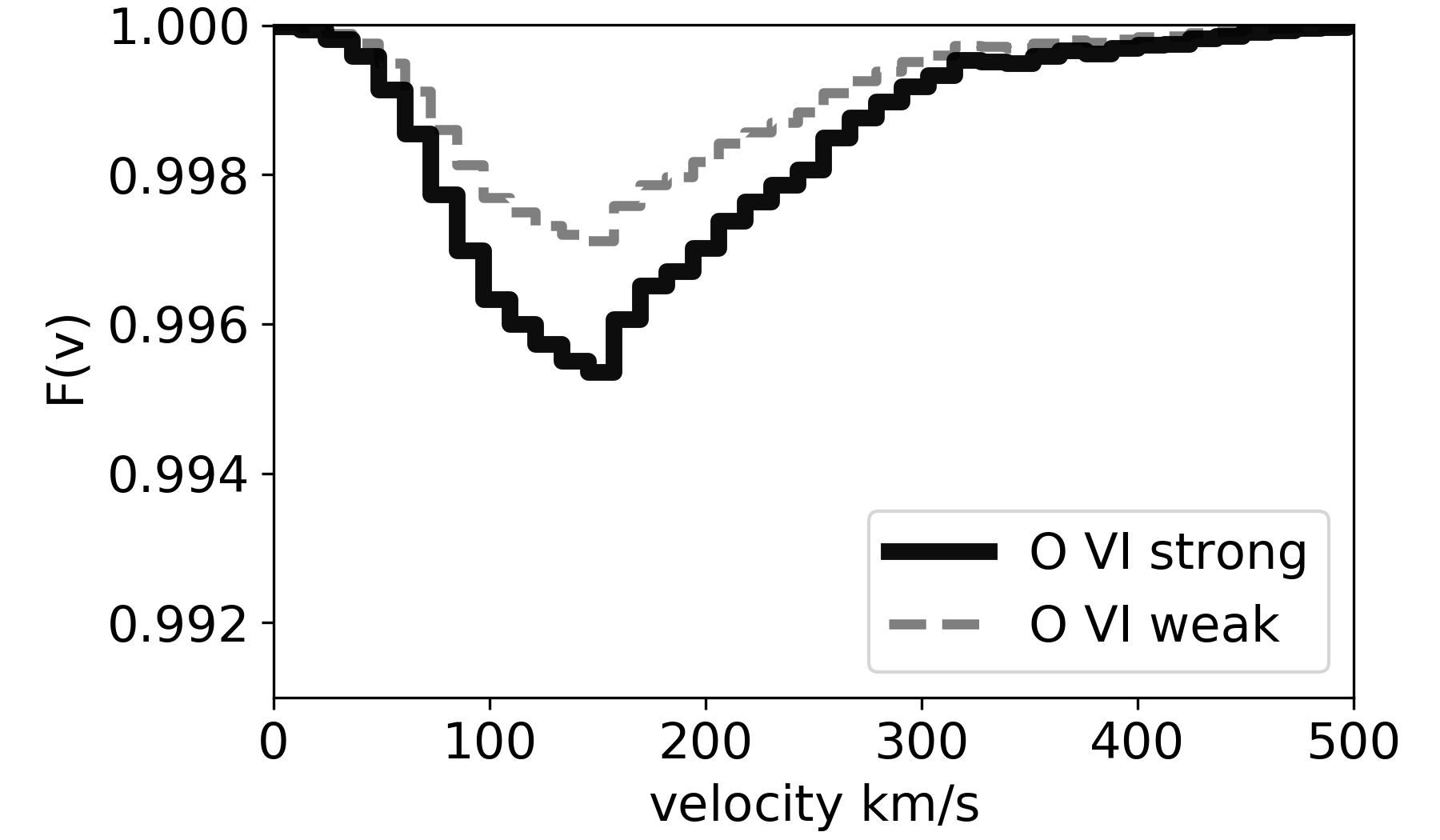}
\caption{Strong and weak absorption lines for the O VI doublet. Although both the strong and weak line peak at the same $v_{\rm min}$ the strong line has a larger $\tau$, and deeper absorption.}
\label{fig: OVI Doublet}
\end{figure}

When the line profiles are optically thin, the depth of the line profiles becomes degenerate with $C_f$ and $\tau$. One observational technique to break this degeneracy and separately compute the optical depth and covering fraction takes advantage of doublets of the same ionic species. For example, Figure \ref{fig: OVI Doublet} shows strong and weak absorption lines for O VI. The strong line for O VI has a larger oscillator strength, $f$, and wavelength coefficients than the weak line. For doublets that have an f-value ratio equal to 2, the relationship between $\tau_{s}$ and $\tau_{w}$ becomes $\tau_{s} (v) = 2 \tau_{w} (v)$. This unique ratio allows one to algebraically solve for the covering fraction independently of $\tau$. In order to facilitate a comparison of our results with observations, we now present ``empirical" estimates for the optical depth and covering fraction, derived according to this method. In all future discussion, ``direct" will refer to the values calculated in Section~3.2, while ``empirical" refers to values calculated using the following observational methodology. 

Following \cite{Chisholm..et..al..2018}, empirical equations were derived for the optical depth and covering fraction of each ionic species. From \autoref{eq:Direct Flux}, if we consider a fraction $C_f(v)$ of the area as covered by a cloud with optical depth $\tau(v)$, and the remaining fraction $1-C_f(v)$ as uncovered, the normalized flux would be
\begin{equation}\label{eq:Normalized Flux}
   F(v) = 1 - C_f(v)+C_f(v) e^{-\tau(v)}. 
\end{equation} 
%where $F(v)$ represents the transmitted flux of each species at any given velocity. 
If $\tau(v)$ for a weak and strong line is related by 
%an $f$-value ratio of 
factor 2, the expression for the weak and strong line can be  can be combined to solve for the covering fraction and the optical depth as a function of the flux in each line, assuming each line has the same $C_{f}$. %The flux for each line can be written as, 
%\begin{equation}\label{eq:Weak and Strong Flux}
%\begin{split}
%F_w(v) = 1 - C_f(v) + C_f(v) e^{-\tau_w(v)},\\
%F_s(v) = 1 - C_f(v) + C_f(v) e^{-\tau_s(v)},
%\end{split}
%\end{equation}
%where $F_{w}(v)$ is the flux for the weak line and $F_{s}(v)$ is the flux for the strong line at any given velocity. By manipulating equation \ref{eq:Weak and Strong Flux} and making these assumption two empirical equations can be derived for the the covering fraction and the optical depth.
The result is
\begin{equation}\label{eq: covering fraction}
     C_f(v)=\frac{(1-F_w(v))^2}{F_s(v)-2F_w(v)+1}
\end{equation}
and
\begin{equation}\label{eq: empirical}
    \tau_0(v)=\ln\bigg[\frac{1-F_w(v)}{F_w(v)-F_s(v)}\bigg],
\end{equation}
where $F_w$ represents the flux from the weak line, $F_s$ is the flux of the strong line, and $\tau_0(v)$ is the optical depth of the  weak line.

 \begin{figure*}[!htb]
\centering
\includegraphics[width=0.3\linewidth]{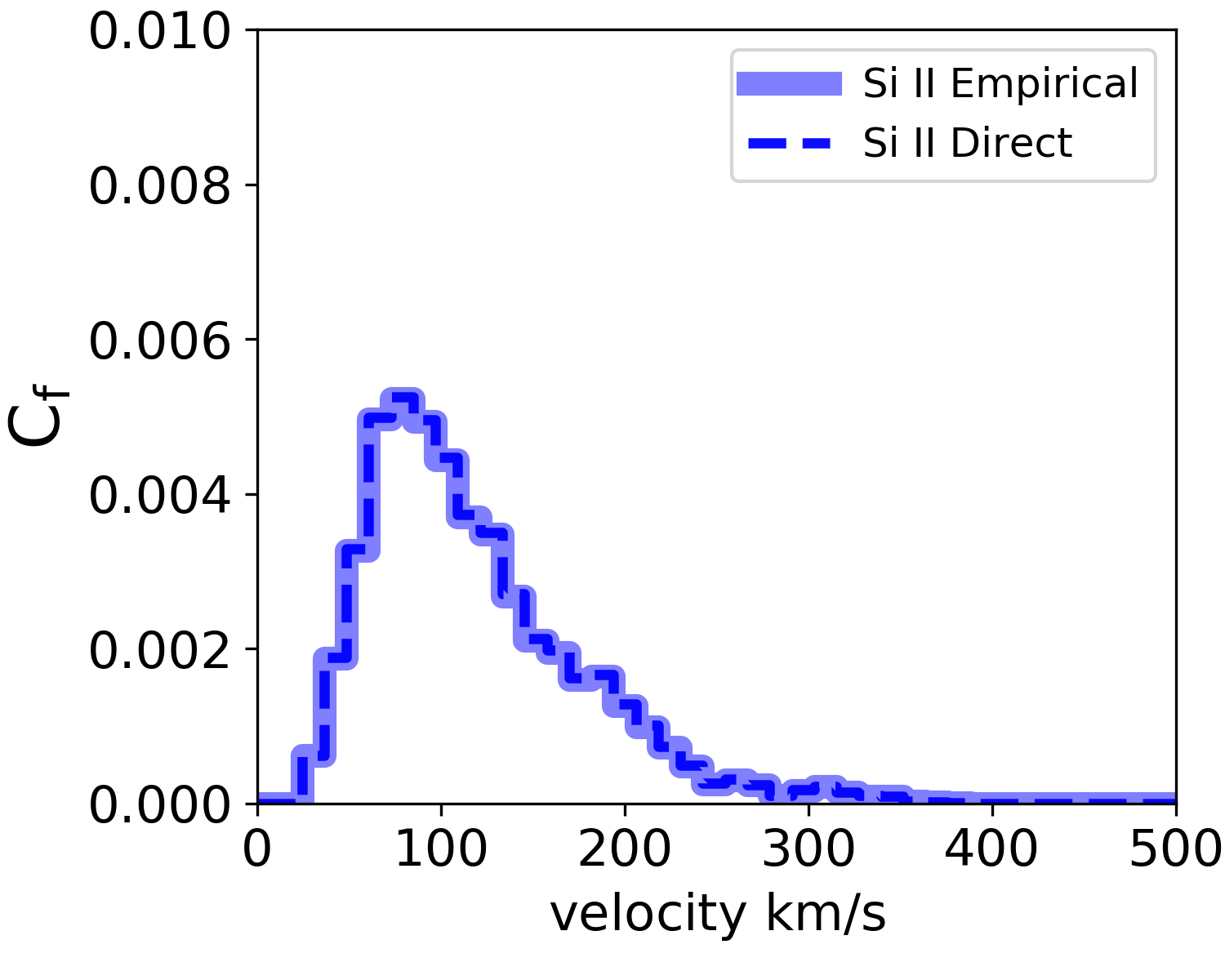}
\includegraphics[width=0.3\linewidth]{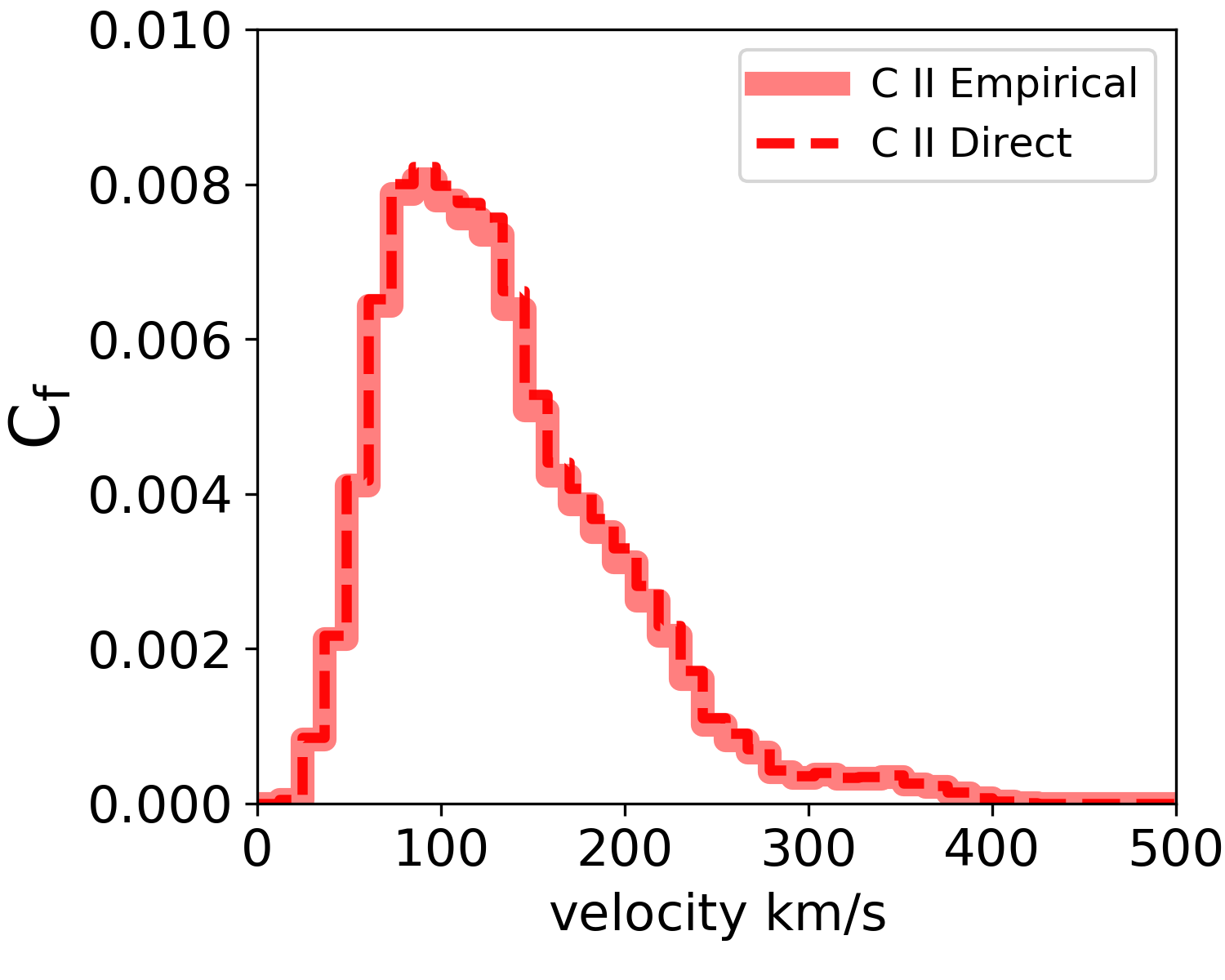}
\includegraphics[width=0.3\linewidth]{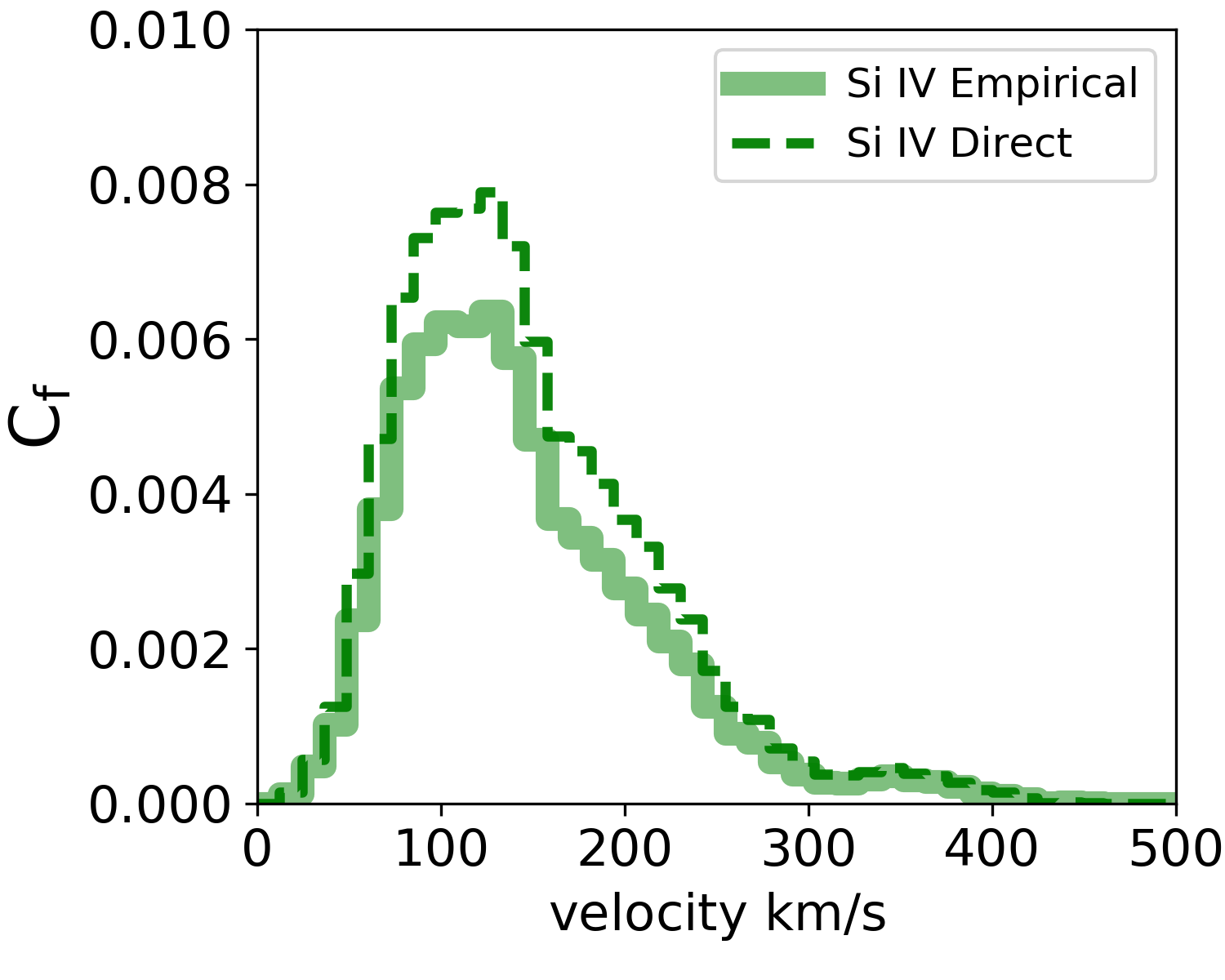}
\includegraphics[width=0.3\linewidth]{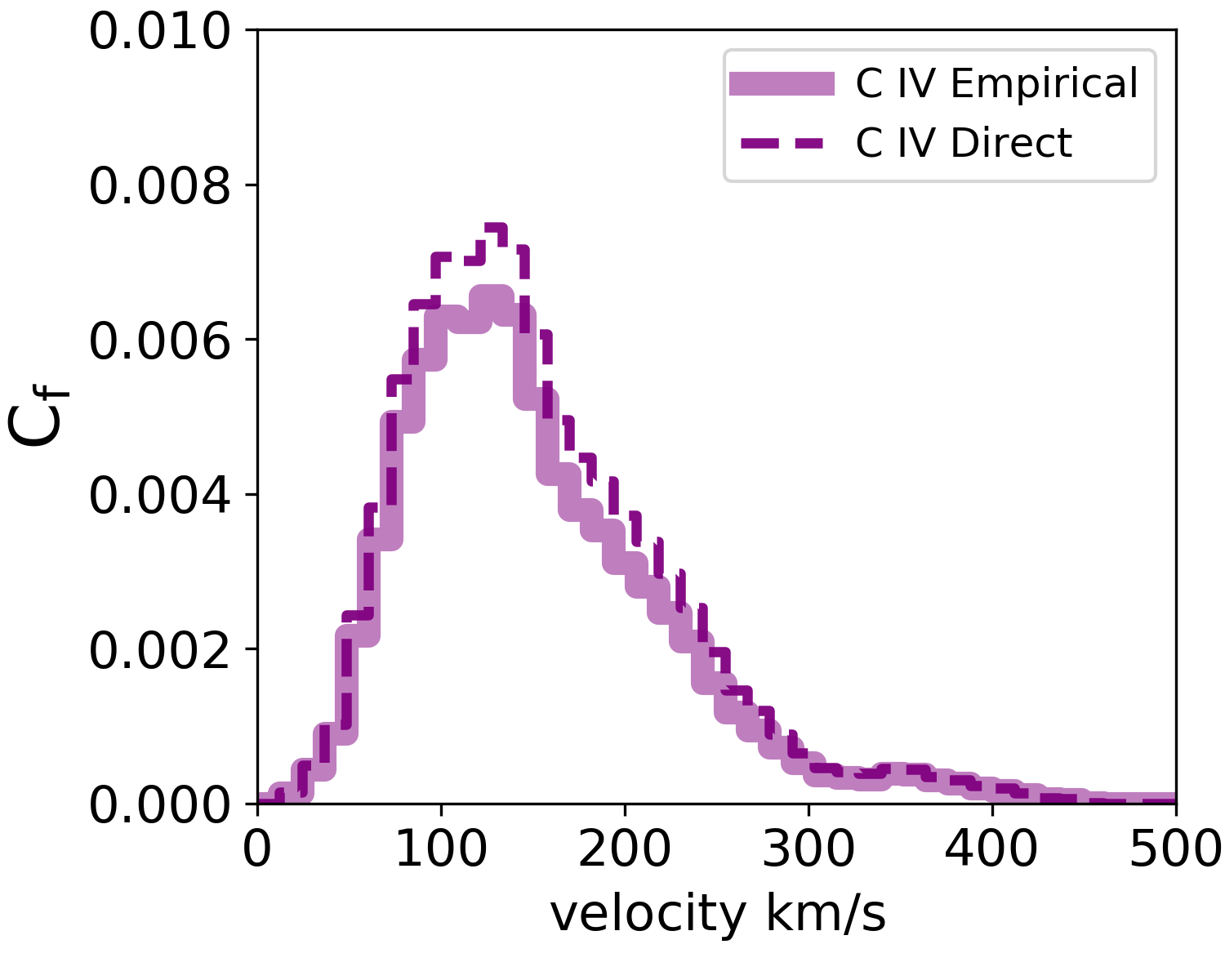}
\includegraphics[width=0.3\linewidth]{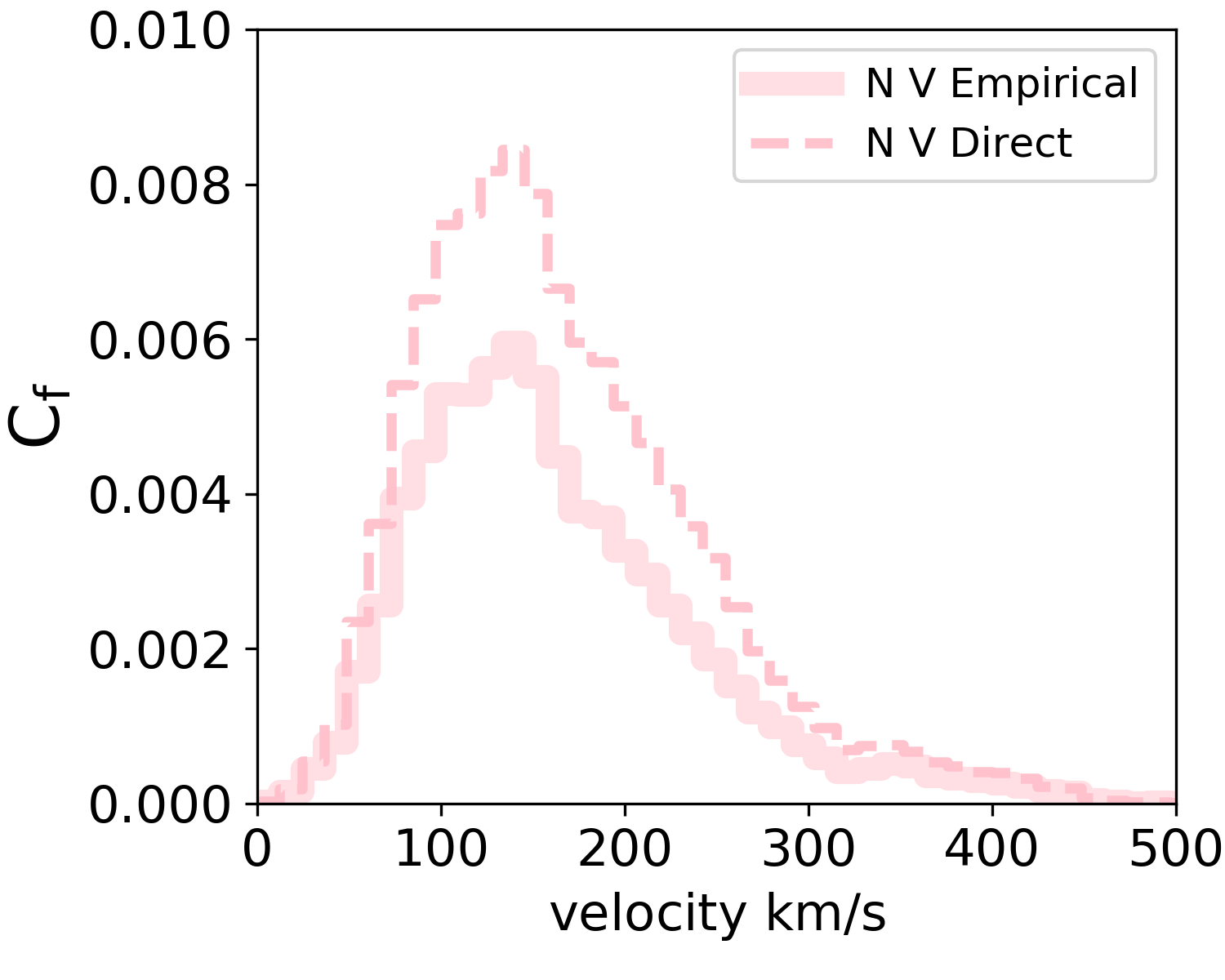}
\includegraphics[width=0.3\linewidth]{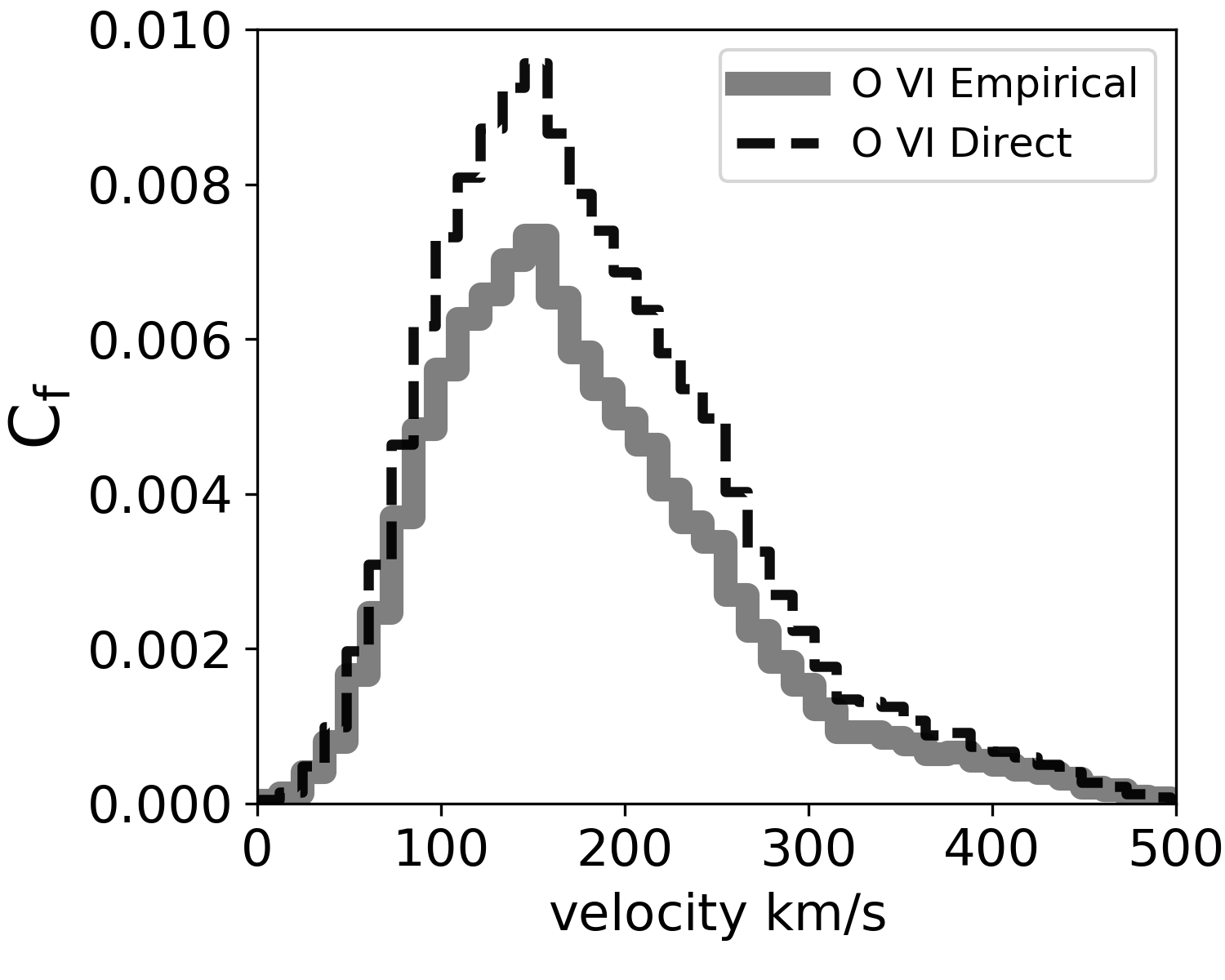}
\caption{A comparison between the empirically derived and direct computation of the covering fraction for strong absorption lines. The empirical and direct method covering fraction calculations agree for low ionization ions, but for the intermediate and high ions the empirical method finds lower values for the covering fraction than the direct method calculation.}
\label{fig: CF}
\end{figure*}

We can use our directly computed flux for both strong and weak lines and  Equation \ref{eq: covering fraction} to calculate an empirical estimate of the covering fraction of each species. In Figure \ref{fig: CF}, we show this empirically calculated covering fraction and compare it to the directly calculated covering fraction, computed by summing all the lines-of-sight with $\tau > 0.1$ and normalizing by the total number, $n_y\times n_z = 262144$ lines of sight along the x-direction. %This comparison between the direct and empirical covering fraction demonstrates that the peak velocity computed directly from our simulation matches the peak velocity obtained using the empirical method. 
There is excellent agreement at all velocities for the low ions, and the modes of the distributions for the direct and empirical method are quite similar for all lines.  
However, the empirically-calculated covering fractions tend to be slightly lower than the directly-computed values for intermediate and high temperature ions.

\begin{figure}
\includegraphics[width=\linewidth]{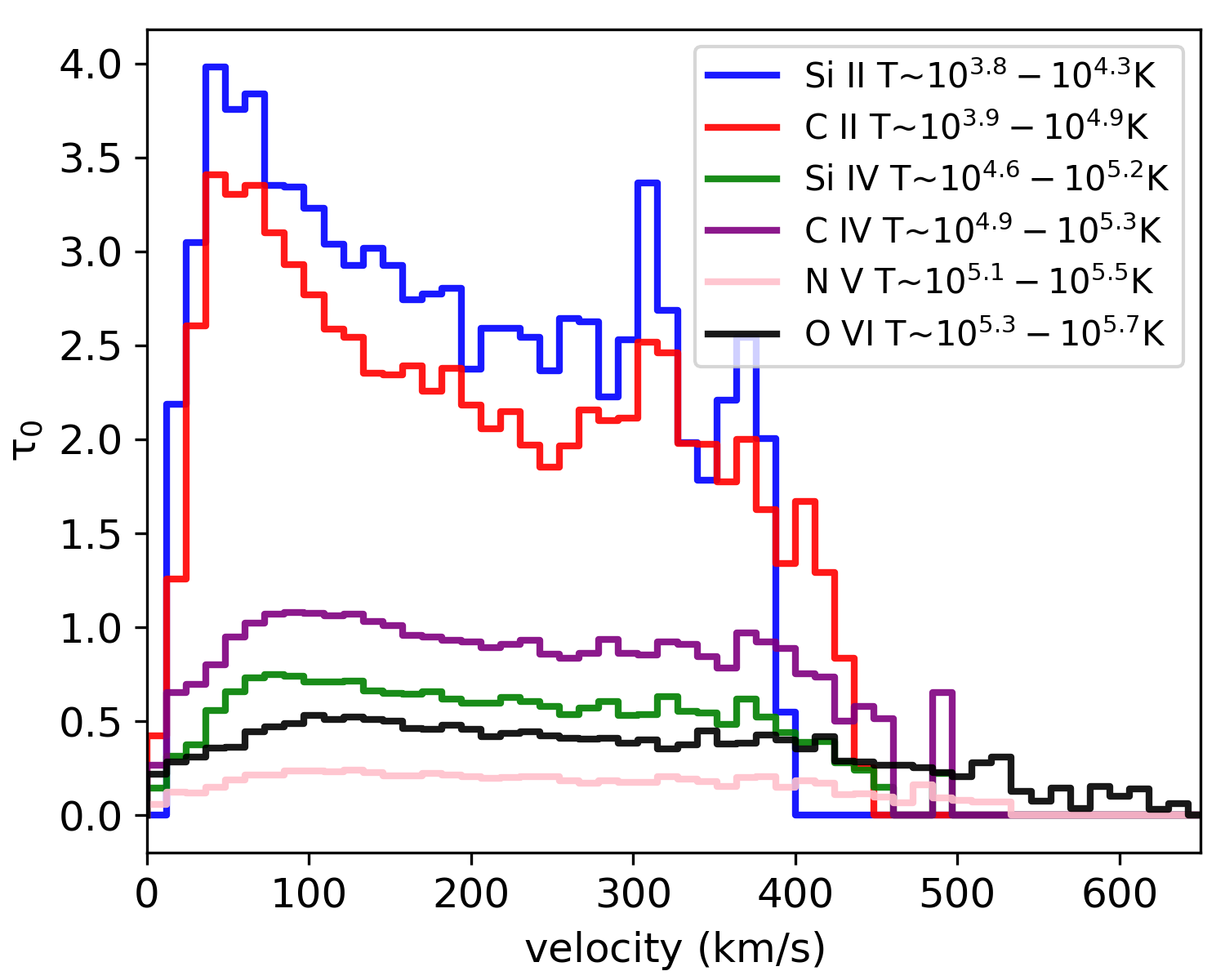}
\caption{The empirical optical depth of doublets of six commonly used ionic species is shown. The lower ions tend to have higher optical depths, and peak at lower velocities.}
\label{fig: AllTau}
\end{figure}
 %Both of the low ionization species, C II and Si II peak at a low velocity of $\sim$ 48.48 km/s
Figure \ref{fig: AllTau} shows the empirically-calculated optical depth, $\tau_0(v)$, for the six different ions that were used for this study. The low ions, Si II and C II, have very similar optical depths. Si II peaks at $\tau_{0}=3.98$ and C II peaks at $\tau_{0}=3.41$; both of these ions also have the same peak velocity of $48$ km/s. This similarity is of course primarily driven by the similar temperature ranges assigned to each ion, and their similar mass fractions under the assumption of solar relative abundances. From this Figure, we also see that the velocity range of the low ions is larger than would be captured by a measurement of $v_{90}$. That is, while most of the mass in these low ions is is at low velocity, there is a tail of high velocity gas even at relatively cool temperatures. The intermediate and higher ions tend to have broader velocity ranges on the whole, and are optically thin across the entire range.
 
\begin{figure*}[!htb]
\centering
\includegraphics[width=0.3\linewidth]{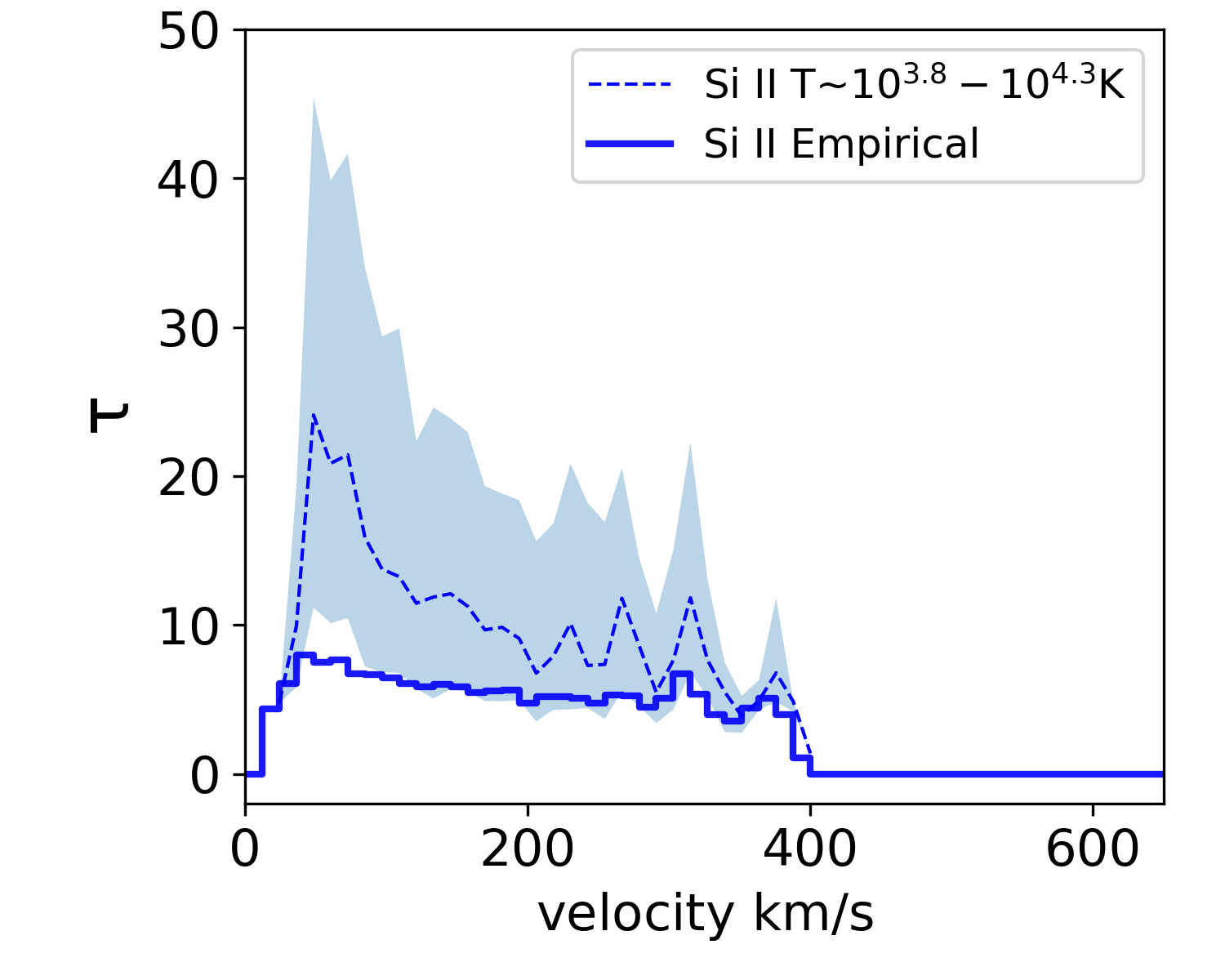}
\includegraphics[width=0.3\linewidth]{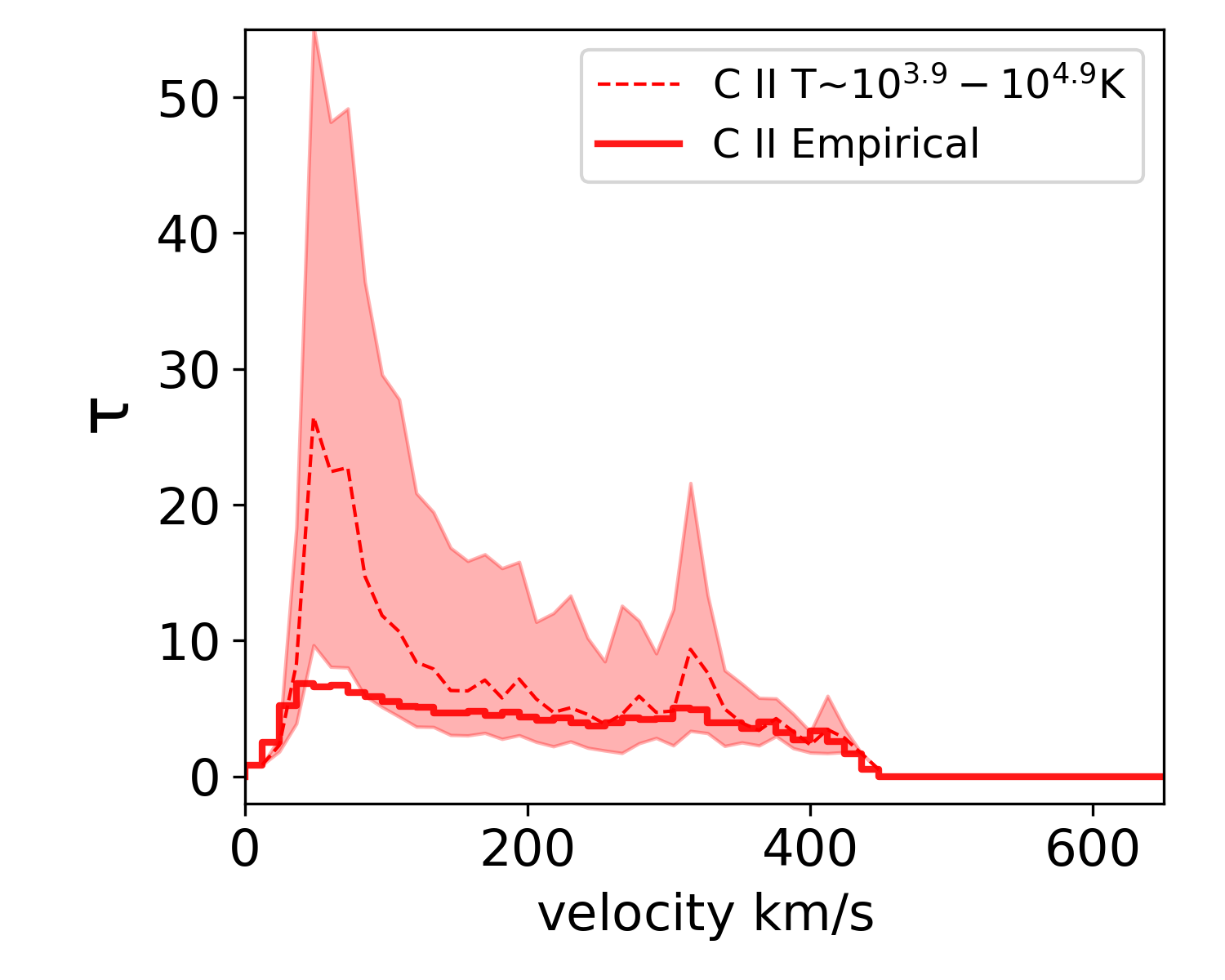}
\includegraphics[width=0.3\linewidth]{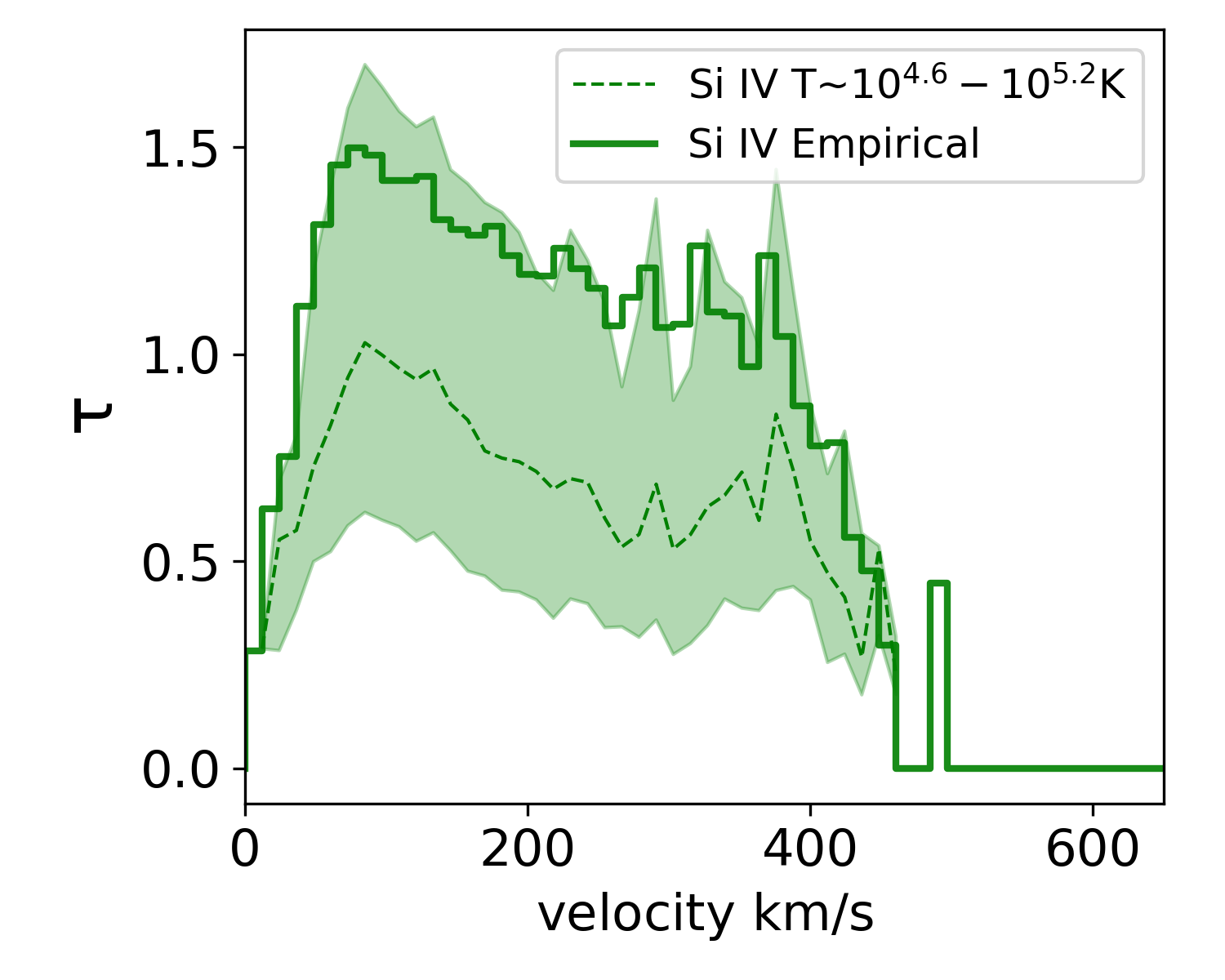}
\includegraphics[width=0.3\linewidth]{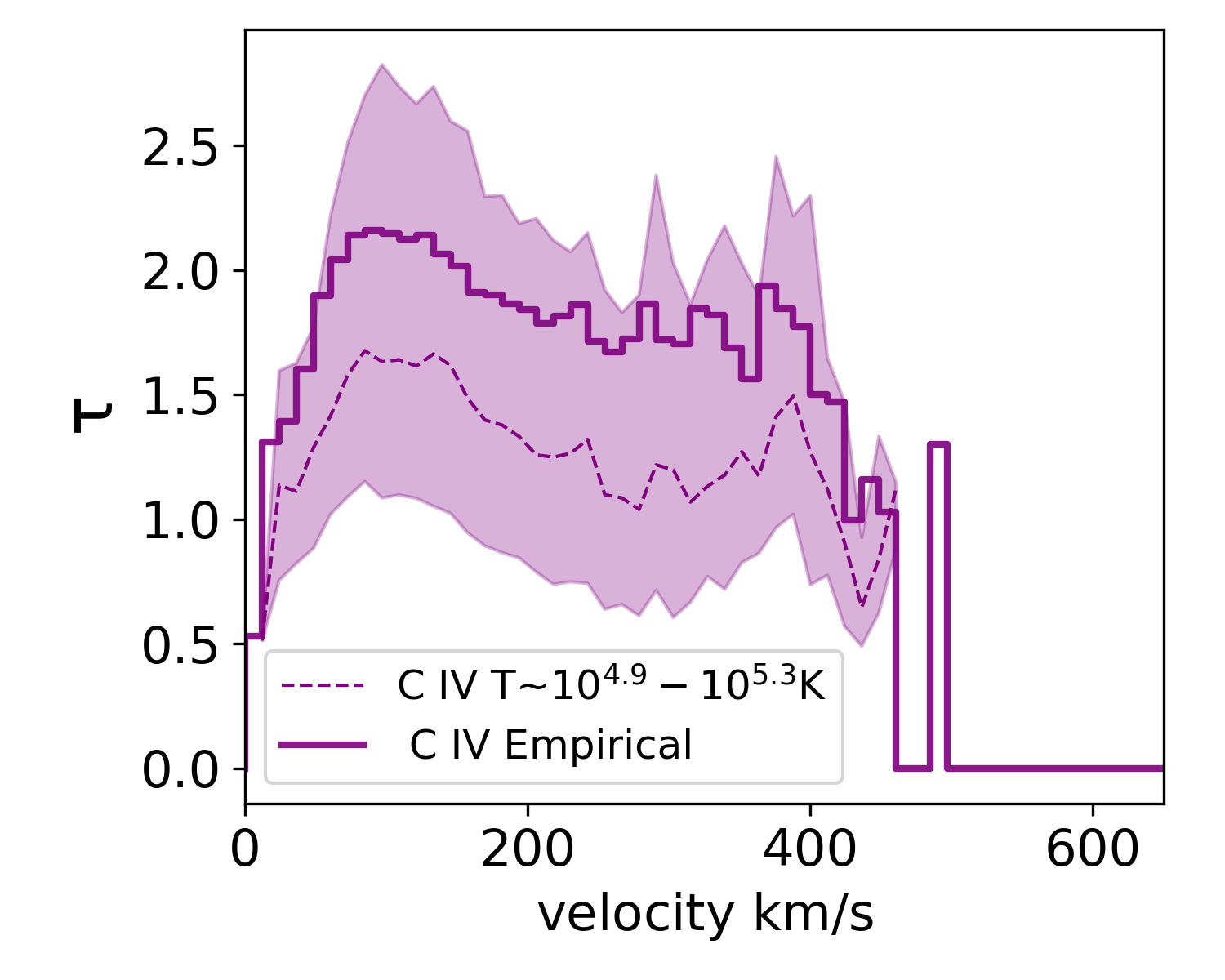}
\includegraphics[width=0.3\linewidth]{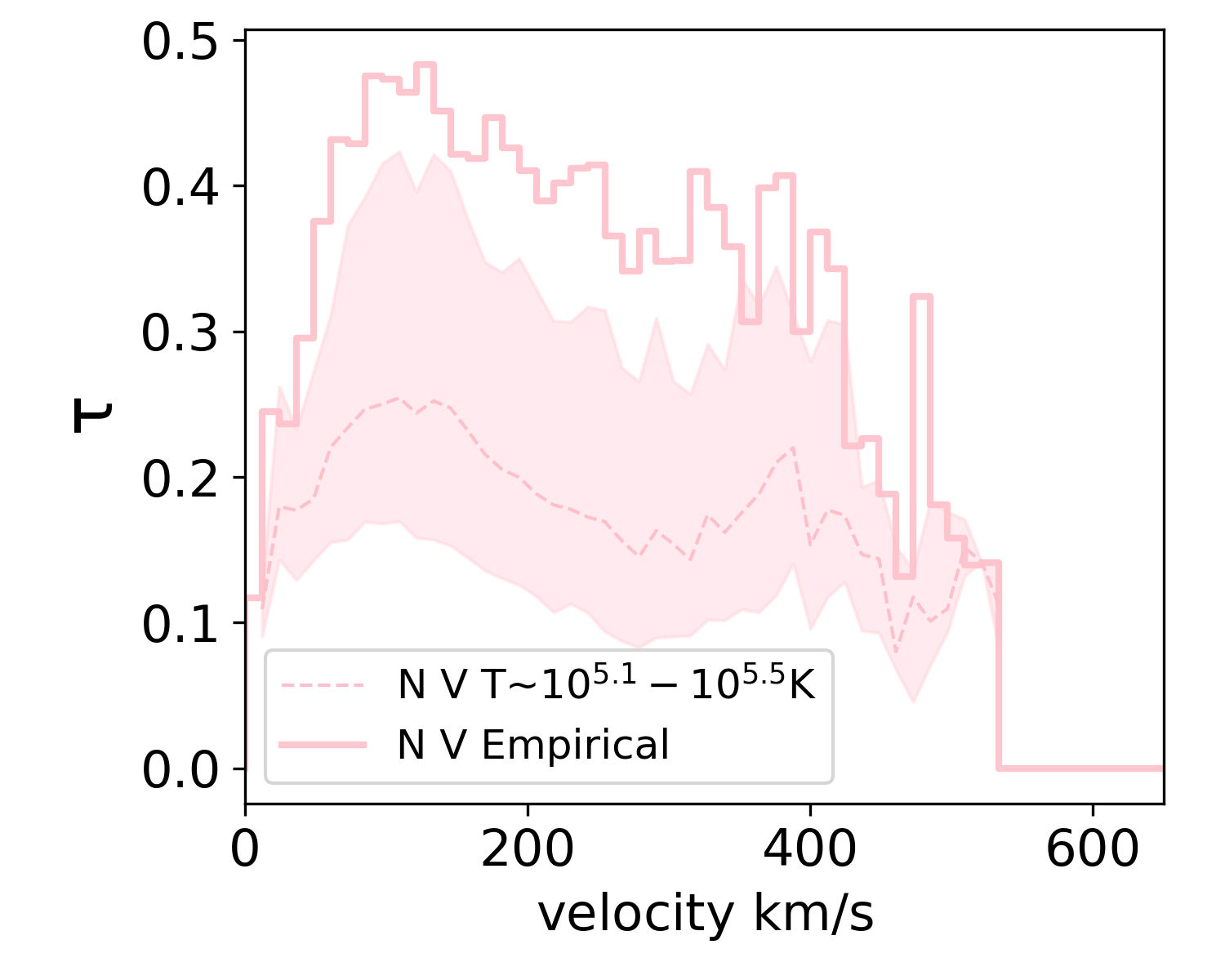}
\includegraphics[width=0.3\linewidth]{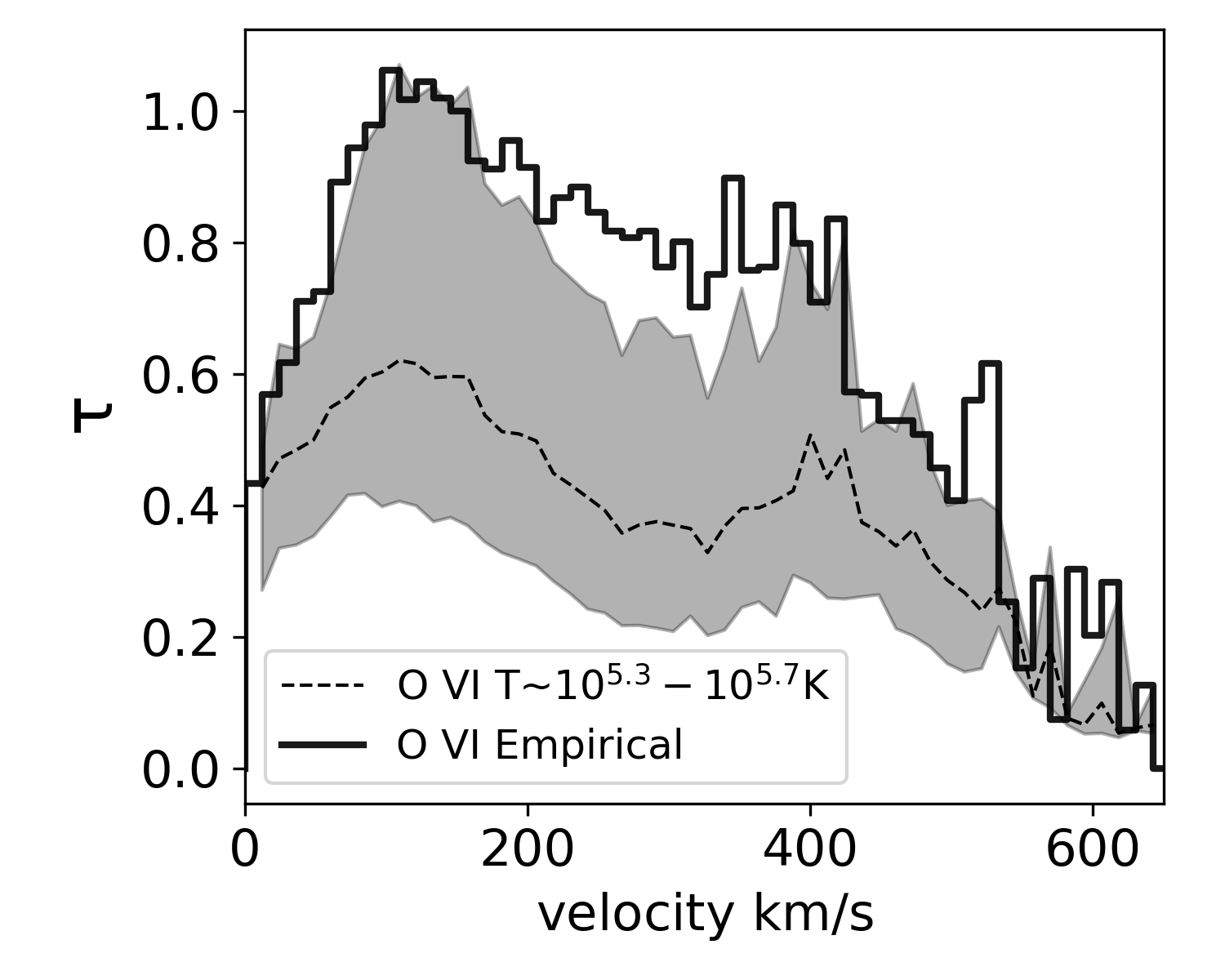}
\caption{An optical depth ($\tau$) comparison between six commonly used species that vary by temperature. The dashed line in each of the panels represents the median value of directly calculated $\tau$. The shaded region represents the 25th and 75th percentile of the directly computed $\tau$ values for all sightlines. The solid colored lines represent the empirically calculated optical depth $\tau_{0}(v)$ for the strong lines.}
\label{fig: Comparison}
\end{figure*}

Figure \ref{fig: Comparison} shows a comparison between the empirical and direct calculations of the optical depth for the weak lines. For the direct method the dashed lines show the median value of $\tau$, while the bottom of the shaded region shows the 25th percentile and the upper boundary represents the 75th percentile. The empirically-calculated values are shown in solid lines. Examining the two upper left panels, we see that low ions tracing low temperature gas tend to have $\tau_{0}(v)$ values that underestimate the directly calculated value of $\tau$. Conversely, in the panels that show our medium and high ions, the value of $\tau_{0}(v)$ consistently overestimates $\tau$. Equation \ref{eq: empirical} can be used to quantitatively explain the cases when $\tau_{0}(v)$ under or over-estimates $\tau$, as follows.

First, we rewrite \autoref{eq: empirical} in terms of the average flux deficit,
\begin{equation}\label{eq: tau0intermsoftauw}
    \tau_0(v)=\ln\bigg[\frac{1-\langle e^{-\tau_w(v)} \rangle}{\langle e^{-\tau_w(v)}(1-e^{-\tau_w(v)})\rangle}\bigg].
\end{equation}
This equation can now be used to describe the two different behaviors of $\tau_{0}(v)$. We define two regimes of $\tau_{0}$ as $\tau_{\rm low}$ and $\tau_{\rm high}$, such that $\langle e^{-\tau_0}\rangle \sim (e^{-\tau_{\rm low}} + e^{-\tau_{\rm high}})/2 $. These values are an arbitrary representation of the lower and upper 50th percentile of $\tau_{0}$. %Allowing $\tau_{w}$ to take a low or high value transforms \autoref{eq: tau0intermsoftauw} into two equations that represent the two regimes of $\tau_{0}$, either optically thin or optically thick. 
In the optically thick case where  $\tau_{\rm high}$ and $\tau_{\rm low}$ $\gg 1$,
\begin{equation}\label{eq: under}
    \tau_0(v)\sim  \ln\left[\frac{2}{e^{-\tau_{\rm high}}+e^{-\tau_{\rm low}}}\right]
    %\rightarrow 
    \sim {\tau_{\rm low}}.
\end{equation}
In the opposite, optically thin regime where $\tau_{high}$ and $\tau_{\rm low}$ $\ll 1$,
\begin{equation}\label{eq: over}
%    \frac{\tau_{low}+\tau_{high}}{e^{-\tau_{high}}\tau_{high}+e^{-\tau_{low}}\tau_{low}} \Rightarrow e^{\tau_{high}}
\tau_0(v) \sim \ln \left[1 + \frac{\tau_{\rm high}^2 + \tau_{\rm low}^2}{\tau_{\rm high}+\tau_{\rm low}} \right]\sim \tau_{\rm high}.
\end{equation}
Equations \ref{eq: under} and \ref{eq: over} can now be used to interpret Figure \ref{fig: Comparison}. Equation \ref{eq: under} shows that when 
%$\tau_{0} >> 1$ it tends to underestimate $\tau$ and when $\tau_{0}<<1$ it overestimates $\tau$. 
optical depth is large, $\tau_0(v)$ tends to lie near the lower range of $\tau$.  Equation \ref{eq: over} shows that when optical depth is small, $\tau_0(v)$  tends to lie near the upper range of $\tau$. 
Figure \ref{fig: Comparison} shows that for the low ions such as Si II and C II that have direct values of $\tau > 1$, $\tau_{0}$ indeed underestimates the true value of $\tau$ (shown by equation \ref{eq: under}). Conversely, for the intermediate and high ionization ions like Si IV, C IV, N V, and O VI, which are optically thin, $\tau_{0}$ overestimates $\tau$.

\section{Discussion}\label{Discussion}

Developing a complete picture of galactic outflows requires a collaborative effort between theorists and observers. In order to interpret the results of numerical models in the context of observed systems, their properties must be translated into physical characteristics that can be observed directly, as demonstrated in Section ~\ref{Results}.  Before embarking on a comparison, however, we should note that the results of our numerical models depend critically on assumptions made about the physical properties of the gas in the simulation. In particular, the predetermined cloud and wind properties will affect the resulting absorption line profiles, covering fractions, etc. The range of cool gas velocities we see in the simulation depends on the initial velocity that is assigned to the hot wind, and potentially its density as well. Similarly, the optical depths and covering fractions values that are found in this study depend on the physical properties assigned to the cloud. Had we chosen a larger , or added multiple clouds along the line of sight, for example, the covering fraction and optical depth values would be larger. Furthermore, the comparison between our results and those from previous work are likely to differ due to different techniques and assumptions that are made. Of particular note is our assumption that all gas is in the wind is in collisional ionization equilibrium. While this is a good starting point, it is likely that cool photoionized gas also contributes to the intermediate ions. This is an avenue we intend to investigate in future work.

\subsection{Observational Comparison}

The ubiquity of outflows and advancements in telescopes and instrumentation in recent years have made it possible to study this multiphase phenomenon with increasingly large samples of objects and with increasingly detailed spectroscopic coverage. Absorption line spectra, in particular, have been used to establish an understanding of the kinematics of galactic outflows as function of different ionic species. In this section, we compare our simulated absorption spectra and kinematics to several representative observational studies. Average characteristics of the galaxies in each study can be found in Table~\ref{table: Galaxy Table}.

\begin{table*}
\caption {Characteristics of Galaxy Samples from the Literature} \label{table: Galaxy Table}
\centering
\begin{tabular}{|c | p{0.25\linewidth} | c | c | c|}
\hline
Reference & Absorption Lines & Sample Size & z & SFR ($M_{\odot} yr^{-1}$ )\\
\hline
\citet{Heckman..et..al..2000} & Na I D & 32 & $<$ 0.1 & 10-100 \\
\citet{Rupke..et..al..2002}   & Na I D & 11 & 0.02-0.27 & 172-612 \\
\citet{Rupke..et..al..2005a, Rupke..et..al..2005b}  & Na I D & 78 &  0.031, 0.129, 0.360 & 40, 225, 389 \\
\citet{Weiner..et..al..2009} & Mg II & 1406 (stack) & 1.4  & 10-50 \\
\citet{Rubin..et..al..2010} & Mg II, Fe II &  468 (stack) & 0.7-1.5 & 1 - 30 \\
\citet{Rubin..et..al..2014} & Mg II, Fe II &  105 &  0.3-1.4 & 0.1-117 \\
\citet{Heckman..et..al..2015} & Si II-IV, Fe II, O I, Al II &  19 of 39 (stack) & $<$ 0.2 & 0.016-66  \\
\citet{Chisholm..et..al..2017} & Si IV & 7 & $\sim0$ & 0.02-26\\
\citet{Chisholm..et..al..2018} & Si I-IV, Fe II, Al II-III, C II, C IV, N V, O I, O VI & 1 & 2.9 & $<$ 48 \footnote{\citet{Wuyts2012}} \\

\hline
\end{tabular}
\end{table*}

We begin our comparison with studies that focused on the Na I ``D" absorption line \citep{Heckman..et..al..2000,Rupke..et..al..2002,Rupke..et..al..2005a, Rupke..et..al..2005b}, which was used to study the warm neutral gas phase in starburst dominated, ultraluminous infrared galaxies (ULIRGs) and infrared galaxies (IRGs). Although we do not attempt to reproduce Na I ``D" because of its sensitive dependence on the dust properties of the  outflow, these studies are nevertheless a good example of early studies with large samples of objects displaying blueshifted absorption.
%so we can use them in comparison to our low ions to compare properties of the low ionization phases.

\citet{Heckman..et..al..2000} investigated 32 far-IR-bright starburst galaxies drawn from parent samples in previous studies \citep{Armus..et..al..1989, Lehnert..et..al..1995}. %The selected objects include strong-stellar-contamination objects (SSC) and interstellar-dominated-objects (ISD). 
The authors found that the absorption line profiles of Na I D in these outflow sources exhibit high maximum velocities that range from 400 to 600 $\rm km \  s^{-1}$. In comparison to our lowest energy ions, we find that our maximum velocities for Si II of 400 $\rm km \ s^{-1}$ and for C II of 448 $\rm km \ s^{-1}$ are within the range of velocities computed in their study, although our $v_{90}$ values tend to be significantly lower.
% The Na I doublet consists of a strong and weak line with an oscillator strength ratio of 2, and thus, it can be used to break the degeneracy between optical depth and covering fraction, as discussed in Section \ref{Optical Depth and Covering Fraction}. Using the ratio of equivalent width of $\lambda$5890 and $\lambda$5896 for Na I D, \citet{Heckman..et..al..2000} computed the central optical depths $\tau_{c}$ in the $\lambda$5896 line for their outflow-dominated objects, finding a range of values from $\tau\sim$ 0.5 to 20, with a median value of 4. We likewise define a $\tau_{c}$ for our simulated data as the $\tau$ obtained after integrating the $\tau$ profiles and identifying the corresponding $\tau$ at 50\% of the area. Comparing our empirically derived $\tau_{c}$ values for strong lines to those in this study, we find that our $\tau_{c}$ values of 4.75 and 4.11 for Si II and C II respectively match the median value well, though half of the \citet{Heckman..et..al..2000} sample thus has higher values. This is not altogether surprising, because Na I D is expected to trace a slightly a lower-temperature, higher density phase than our Si II and C II profiles.

Using the ESI instrument on Keck II, \citet{Rupke..et..al..2002} studied 11 ULIRG from the IRAS 1 Jy survey that have visible Na I D absorption.  The authors measured a ``typical" outflow speed of 300 $\rm km \ s^{-1}$, which is consistent with our low ion values for $v_{90}$, but significantly higher than our values for $v_{central}$.
%Most of the $\tau_{c}$ values in this study for the strong lines of the Na I D doublet have relatively low $\tau_{c}$ ($<$2)
%The $\tau_{c}$ values in this study range from 0.12 to 7.5 with only four values being 2.4, 3.44, 6.28, and 7.5 putting our calculated values in the higher end of this spectrum. 
%putting our empirically derived central optical depth values of 4.75 and 4.11 for Si II and C II at the higher end of this spectrum. %To fit multiple absorbing components and Gaussians in optical depth this study used an IRAF-based interactive spectral fitting software (SPECFIT). This algorithm 
%The authors note that differences in fitting algorithms could be responsible for the smaller values of $\tau_{c}$ found in this study versus the values presented by \cite{Heckman..et..al..2000}. Regardless, the values found in both studies are consistent with our results for the lowest ionization state gas. %Furthermore, the physical properties of the sample used in each study ULIGs \citep{Rupke..et..al..2002} and low-luminosity galaxies \citep{Heckman..et..al..2000} could also be a contributing factor to the discrepancies between both values of $\tau_{c}$ calculated in each study.
\citet{Rupke..et..al..2005a, Rupke..et..al..2005b} studied 78 ULIRGs and IR-galaxies at $z < 0.5$.  Because of the large sample size, the authors were able to study the evolution of winds as a function of host galaxy properties. The sample was split into three different categories: infrared galaxies (IRGs) (z $<$ 0.5), low-z ULIRGs (z $<$ 0.25), and high-z ULIRGs (0.25 $<$ z $<$ 0.50). ULIRGs, $L_{\rm IR} > 10^{12} L_{\odot}$ are host to starbursts and power star formation with the large amount of dense molecular gas found in their cores. By contrast, IRGs ( $10^{10} L_{\odot} < L_{\rm IR} < 10^{12} L_{\odot}$) host fewer starbursts and have a lower wind detection rate than ULIRGs \citep{Heckman..et..al..2000, Rupke..et..al..2002, Martin..2005}. Each of the three samples were selected from three different surveys; the IRGs from the the Infrared Astronomical Satellite (IRAS) Revised Bright Galaxy Sample (RBGS) and Warm Galaxy Sample (WGS), the low-z ULIRGs from the IRAS 1 Jy survey, and the high-z ULIRGs from the FIRST/FSC survey. 
%From \citet{Rupke..et..al..2005a} we compare the empirical central optical depth for Si II and C II with those computed for each of the three different galaxy samples and find that our results are within the range for IRGs ($\sim \tau$ 0.07 to 5.00) and low-z ULIRGs ($\sim \tau$ 0.06 to 5.00), but above the range computed for high-z ULIRGs($\sim \tau$  0.09 to 2.00). 
We compare our maximum velocities for Si II and C II with the average maximum velocities for the three samples, which are 308, 401, and 359 $\rm km \ s^{-1}$ for IRGs, low-z ULIRGs, and high-z ULIRGs  respectively.  Our maximum velocity for Si II (400 $\rm km \ s^{-1}$) closely matches that of the low-z ULIRGs; while our maximum velocity for C II (448 $\rm km \ s^{-1}$) exceeds the average maximum velocity computed in this study for all three samples. 

Starting with stacks of galaxies at intermediate redshift, and later individual galaxies at lower redshifts, \citet{Weiner..et..al..2009,Rubin..et..al..2010,Rubin..et..al..2014,Heckman..et..al..2015} used starburst galaxies to study UV absorption lines of the warm ionized gas ($10^{4}-10^{5}\,\rm K$) in outflows.
%Mg II and Fe II with an ionization potential of 15.0 $\rm eV$ and 16.18 $\rm eV$ respectively can be compared to our low and intermediate temperature ions Si II (16.35 $\rm eV$) , C II (24.38 $\rm eV$), Si IV (45.14 $\rm eV$), and C IV (64.49 $\rm eV$).
\citet{Weiner..et..al..2009} studied Mg II absorption in the stacked spectra of 1,406 starburst galaxies at z $\sim$ 1.4 from the DEEP2 redshift survey. With an ionization potential of 15.0 $\rm eV$, Mg II does not photodissociate as easily as Na I D which leads to smaller ionization corrections. This study found central outflow velocities that range from 250 to 300 $\rm km \ s^{-1}$. Our $v_{\rm central}$ for Si II, C II, Si IV, and C IV are 132, 144, 156, 156 $\rm km \ s^{-1}$ respectively, which are consistently smaller than the velocities found in this study. In a similar study, \citet{Rubin..et..al..2010} investigated Mg II and Fe II absorption in the stacked spectra of 468 galaxies from the Team Keck Treasury Redshift Survey (TKRS) at slightly lower redshift.
%This study finds $\tau_{c}$ for the strong lines of Mg II is $\tau_{c, \rm MgII} \sim \ 3$ and for Fe II is $\tau_{c, \rm FeII} \sim \ 3$ as well. These values are very similar to our $\tau_{c}$ values for Si II and C II (4.75 and 4.11), and slightly higher than our values for the higher energy ions Si IV and C IV (1.4, and 1.71).
%Using the velocity of the O II emission line doublet the authors calculated the systemic velocity given the redshift of the galaxy. 
The authors found central velocity offsets of $\sim 250\,\rm km \ s^{-1}$ for Fe II and Mg II, with slightly larger offsets for Mg than Fe. These values are again slightly larger than our $v\mathrm{central}$ values for the low ions.

\citet{Rubin..et..al..2014} used redshift surveys of the Great Observatories Origins Deep Survey (GOODS) fields and the Extended Groth Strip (EGS) to analyze the individual spectra of 105 galaxies using the Mg II ($\lambda$2796 and $\lambda$2803) and Fe II ($\lambda$2586 and $\lambda$2600) doublet absorption line profiles at 0.3 $<$ z $<$ 1.4. Two distinct models were used in this study in order to quantify the kinematics and strength of each line profile. The ``one-component" model follows the same approach as \citet{Rupke..et..al..2005a} and parameterizes the normalized flux as a function of wavelength. The ``two-component" model assumes that the same line profile comes from two velocities as opposed to one. We compare our $v_\mathrm{central}$ with the two-component model, for which most of the calculated $v_\mathrm{central}$ values are $> 150 \rm km \ s^{-1}$, and find that our values for Si II, C II, Si IV, and C IV of 132, 144, 156, 156 $\rm km \ s^{-1}$ respectively are within the range of velocity values calculated in this study. 

%the range of values of both the one-component and two-component model of Mg II found in the study, which range between 25 and -403 $\rm km \ s^{-1}$.  Comparing our $v_\mathrm{central}$ we find that are values are within the computed velocities of the Fe II absorption one and two component models with ranging velocities of 4-236 $\rm km \ s^{-1}$ and 3-429 $\rm km \ s^{-1}$ respectively. 

\citet{Heckman..et..al..2015} studied the kinematics of the warm gas phase found in supernovae driven winds of 39 low-redshift (z $<$ 0.2) objects using  UV absorption lines. This study used two samples of galaxies, the stacked spectra of 19 galaxies observed with the Far Ultraviolet Spectroscopic Explorer satellite (FUSE); and the individual spectra of 21 Lyman Break Analogs (LBAs) observed with COS on HST. We compare our absorption line profiles of Si II, C II, and Si IV with those computed in this study and find that our $v_{min}$ values of 84, 96, 133 $\rm km \ s^{-1}$ are systematically smaller than those found in this study. The $v_{min}$ values that \citet{Heckman..et..al..2015} find for the same species (Si II, C II, and Si IV) are 200, 150, 300 $\rm km \ s^{-1}$ respectively.

Using COS spectra \cite{Chisholm..et..al..2017} investigated 7 starburst galaxies using UV absorption lines. Using the Si IV doublet they derive and solve a system of two equations for the velocity-resolved optical depth and covering fraction, as described in Section~\ref{Results}. We compare our optical depth and covering fraction results for Si IV to those from this study. The peak $\tau$ values for Si IV in this study range from $\tau \sim$ 0.5-2.7, comparable to our $\tau_0$ value for Si IV which peaks at 0.75. A notable difference between our work and this study is the assumption in \cite{Chisholm..et..al..2017} that the Si IV absorption is produced exclusively by cool photoionized gas, while we assume that it is produced exclusively by warm collisionally ionized gas. Our covering fractions are of course much smaller than those computed in this study due to the small area of the cloud in our simulation.

In a particularly instructive single-object study, \cite{Chisholm..et..al..2018} used data from the the Magellan Evolution of Galaxies Spectro-scopic and Ultraviolet Reference Atlas (MEGaSaURA) to study O VI 1032$\mathring{A}$ absorption from a gravitationally lensed high-redshift galaxy (z $\sim \ 2.9$) with a star formation rate of $<48\ M_{\odot} \rm yr^{-1}$ as measured by SED fit.
O VI in winds likely probes gas transitioning between a warm $10^4$K gas and a hot $>10^7 \rm K$, allowing the kinematics of distinct gas phases to be studied. This study examined absorption line profiles of many lower ions (see Table~\ref{table: Galaxy Table}) and compared them to O VI, finding two ``regimes" for the OVI line - at low velocities, its profile resembles that of the optically-thick low ions, while at high velocities its profile resembles that of the optically-thin lines. In general, we find that the optical depth of our lines is smaller than those shown in their work. This is partly due to the fact that the depth of our normalized flux is affected by the size of our simulation and the size of our cloud. When calculating our flux we average over the entire simulation box, so there are many unobstructed sight lines in our average, leading to shallow absorption lines. However, we can more fairly compare the quantities $v_{\rm central}$ and $v_{90}$ from the studied metal absorption lines. In contrast to the previous studies mentioned, when comparing our computed velocity values to the derived quantities from \cite{Chisholm..et..al..2018} we find that our values do not match the observations well. In this study the central velocities for the strong lines of Si II, C IV, and O VI are 201, 130, and 246 $\rm km \ s^{-1}$ while our values are systematically lower, at 132, 156, and 180 $\rm km \ s^{-1}$ respectively. Our $v_{90}$ values for the same three species (228, 264, and 300 $\rm km \ s^{-1}$) also underpredict those seen in this object - \citet{Chisholm..et..al..2018} find values corresponding to these same species of 585, 455, and 532 $\rm km \ s^{-1}$.

Generally speaking, our lowest ionization state gas appears to match the kinematics of the low-z observed absorption-line systems better than the high-z systems. Our numerical study is based on the canonical low-z starburst galaxy M82, and assumes a lower star formation rate (and thus less extreme hot wind properties) than many of the high-z systems discussed here, which may account for the better match to the systems observed a lower redshift. On the other hand, our higher ionization lines consistently underpredict the observed values for all studies. In particular, our higher ionization lines tend to have lower central and maximum velocities than those that are observed. In a number of recent studies, several authors have demonstrated that additional momentum can be transferred to gas in the warm photoionized phase and intermediate hotter phases via mixing with the hot wind, and that the amount of momentum transferred is a function of the amount of mixing that has occurred, and thus the distance that the gas has traveled \citep[e.g.][]{Gronke..et..al..2018, Gronke2020, Fielding2018, Vijayan2020, Schneider2020}. Because the simulation used in this study consisted of a small (5 pc) cloud in a relatively small (160 pc) box, there is simply not enough room for the higher ionization state gas to have achieved the velocities commonly observed. In larger scale simulations, this gas experiences further acceleration, and the higher ionization state profiles may be a better fit \citep{Schneider2020}. This is an avenue we will explore in future work.

\subsection{Simulations}

In addition to the many observational studies, the past couple of years has seen a substantial effort on the theoretical side to better understand the physics of outflows through the use numerical simulations \citep[e.g][]{Cooper..et..al..2009, Scannapieco..et..al..2015, Bruggen..et..al..2016, Schneider..et..al..2017, Gronke..et..al..2018, Cottle..et..al..2018, Kim..et..al..2018, Fielding2018, Kim..et..al..2020}. Most of these studies did not include absorption spectra and we therefore do not compare the results here, however, we can compare to \citet{Cottle..et..al..2018}, who included a thorough examination of blue-shifted absorption from an idealized wind simulation very similar to that presented in this work.
 
\citet{Cottle..et..al..2018} used adaptive mesh refinement 3D hydrodynamic simulations from \citet{Scannapieco..et..al..2015} and \citet{Bruggen..et..al..2016} to study a cool cloud in a hot wind. These simulations included the effects of radiative cooling, thermal conduction, and an ionizing background characteristic of a starburst galaxy in varying combinations. The authors used the TRIDENT code \citep{Hummels..et..al..2017} to produce synthetic column densities of 10 different species from the simulations: H I, Mg II, C II, C III, C IV, Si III, Si IV, N V, O VI, and Ne VIII. The physical volume of the simulation box hosting the interaction between these two phases is -800 $\times$ 800 $\rm pc$ along the x-y direction and -400 $\times$ 800 $\rm pc$ along the z direction which is the direction of the hot wind. The simulations used a cloud with an initial temperature of T = $10^{4} \rm K$ , radius of $R_{\rm cloud} = \ 100 \rm pc$ , and a mass density of $\rho$ = $10^{-24} \rm  g \ cm^{-3}$ (substantially larger than the cloud in this work), and allowed cooling down to $10^4\,\rm{K}$. The study included results at four characteristic times in the cloud's destruction that correspond to when 95\%, 75\%, 50\%, and 25\% of the mass fraction of the cloud remains at or above 1/3 of the cloud’s original density. Here we compare our results to their $t_{50}$ which is the most similar to our 10 $t_{\rm cc}$ snapshots.

We compare our column densities of C IV and O VI to those computed in this study and find that similar to our findings O VI tends to be more spatially extended than the lower ions. As in our simulation, the column density profiles of all the ions in \citet{Cottle..et..al..2018} trace each other fairly well indicating that the species are spatially related and have a similar creation mechanism. Comparing to our normalized flux results, we find similar results only for N V - all other ions in our study have much smaller depths than theirs, and their lowest ionization states are completely saturated. This can primarily be attributed to the much larger cloud used in the \citet{Cottle..et..al..2018} work, which leads to larger column densities, optical depths, and covering fractions. When comparing our velocities with those in this study we see that for both our model and \citet{Cottle..et..al..2018} the velocities increase with ionization state. This stands in contrast to the velocities shown in \citet{Chisholm..et..al..2018} Table 1, where there is not a clear correlation between increasing ionization states and velocities. Additionally, we compare our covering fraction results to those from this study and find that our values are smaller than those presented in this study. Although our velocity profiles do not have a similar shape to those presented in \citet{Cottle..et..al..2018}, we do find similar $v_{\rm min}$ values - our velocities all fall within their peak range of $\sim$ 50 to 200 $\rm km \ s^{-1}$ for the various ions.

\section{Conclusions}\label{Conclusion}

In this work, we have presented a study of the gas in multiphase outflows using a high resolution numerical simulation that models the interaction between a hot supernovae driven wind and cool cloud of interstellar material \citep{Schneider..et..al..2017}. Using this simulated dataset, we have created observables that we compare to observational studies of outflows, in order to better understand the kinematics and phase structure of the gas. Gas in the simulation has a wide range of temperatures, densities, and velocities. By assigning ionization states to the gas using collisionally ionized gas temperatures, we enable a direct comparison to the individual species observed in absorption line studies of outflows.

In particular, we create synthetic absorption lines for a sample of representative ions, which are then used to calculate the optical depth and covering fraction of individual species. These absorption lines are characterized by a few general features: as  ionization state increases so do velocity centroids and maximum values (see Table ~\ref{table: velocity table}). Optical depths are higher for the lower ions, and covering fractions are lower, although marginally. All of our covering fractions are very low compared to the observations, which is a result of our small cloud size compared to the simulation domain. Consequently, the absorption as shown by the normalized fluxes presented in this work is very shallow relative to observed lines. We therefore focus primarily on the shapes of the velocity profiles and associated statistics. %Furthermore, we notice that our low ionization lines tend to peak at lower velocities that our high ionization lines. Looking at our velocity values for $v_{\rm min}$, $v_{\rm central}$, $v_{90}$ in Table 2 we find that these velocities for our intermediate ions (Si IV and C IV) are exactly the same this is representative of the fact that these ions are spatially related. 

Motivated by \cite{Chisholm..et..al..2018} we also derived an ``empirical''  optical depth and covering fraction using doublets from the same ionic species. We compared these values with the directly computed optical depth and covering fraction values obtained using the simulation. Figure \ref{fig: Comparison} shows that the empirically derived optical depth for the optically-thick low ionization ions like Si II and C II tends to underestimate the true values of $\tau$. Conversely, the empirical optical depth values for the intermediate and high ions overestimates the true value of $\tau$. It is nevertheless true that the ``empirical'' optical depth is still between the 25th and 75th percentile of the optical depth distribution for most ions. We demonstrate that biases in optical depth estimates can be understood based on whether the absorption lines are optically thick or optically thin (see Eqns \ref{eq: under} and \ref{eq: over}).   

In Section~\ref{Discussion}, we compare our results with an extensive sample of observational studies from the literature that investigate line profiles of species with similar ionization states (e.g Na I D, Mg II, and Fe II). We find that that our maximum velocity for Si II and C II resembles the velocities reached by Na I D. Similarly, our central velocities for Si II, C II, Si IV, and C IV are within the typical velocity ranges of most UV absorption line studies, particularly for lower star formation rate systems more comparable to our assumed wind. Our higher ionization lines undershoot the maximum velocities reported in most systems, which is a result of our small simulation box size and limited physical range for acceleration.

This study is an first step towards more detailed comparisons of simulated and observed absorption line data. In future work, we hope to expand the results of this work to larger simulations, and improve the physical accuracy of the simulated absorption lines through full radiative-transfer modeling. Nevertheless, we anticipate that the results presented here will provide additional insight into the nature of the multiphase gas in galactic outflows, and how it gives rise to the observed fluxes. These comparisons between simulated data and observations will ultimately help improve both our theoretical models and our ability to interpret the kinematics of galactic outflows.
%Furthermore we compare our results with other observational that studies that study similar species (e.g Na I D, Mg II, and Fe II). We find that that our velocities for low ionization ions such as Si II and C II closely match the maximum velocity reached by Na I D. In addition, we also compare our central velocities with UV absorption line studies and find that our velocities for Si II, C II, Si IV, and C IV are within the velocity range provided by these studies. When comparing the minimum velocity with UV absorption line studies we find that our velocities are smaller than those computed in those studies. These comparisons are done with ions that have similar ionization states as the ions we have pre-selected based on having an f-value ratio of 2.

%In addition, to comparing our work with observations with also compare our work with a hydrodynamical simulation code carried out with FLASH that models a cool cloud driven by a hot supersonic wind. We find that our covering for our high ions O VI and N V match very closely to those computed using this code. We also find that our column density profile for our high temperature ion O VI close matches the one traced out in this study. For our low and intermediate temperature ions we find that our absorption line profiles don't correlate with those from this study ours are not as steep.

\section{Acknowledgments}
This research used resources of the Oak Ridge Leadership Computing Facility, which is a DOE Office of Science User Facility supported under Contract DE-AC05-00OR22725, via an award of computing time on the Titan system provided by the INCITE program (AST125). E.E.S was supported in part by NASA through Hubble Fellowship grant \#HF-51397.001-A awarded by the Space Telescope Science Institute, which is operated by the Association of Universities for Research in Astronomy, Inc., for NASA, under contract NAS 5-26555. The work of ECO was partly supported by grant 510940 from the Simons Foundation.

\software{Cholla} \citep{Schneider..et..al..2015}; \texttt{matplotlib} \citep{Hunter2007}, \texttt{numpy} \citep{Vanderwalt..et..al..2011}, \texttt{hdf5}

%\appendix

%\section{Appendix material}
\newpage

\bibliography{References}

\end{document}